\documentclass[fleqn,usenatbib,useAMS]{mnras}

\usepackage{graphicx}	
\usepackage{amsmath}	
\usepackage{amssymb}	
\usepackage{multicol}        
\usepackage{bm}		
\usepackage{pdflscape}	
\usepackage{multirow}
\usepackage[outdir=./]{epstopdf}
\usepackage{ulem}
\usepackage{hyperref}
\usepackage[flushleft]{threeparttable}

\usepackage[dvipsnames]{xcolor}
\usepackage{xparse,xcolor}

\def\Msun{{\rm M_{\odot}}}
\def\Rsun{{\rm R_{\odot}}}
\def\Lsun{{\rm L_{\odot}}}

\def\Msuny{{\rm M_{\odot}}{\rm yr^{-1}}}
\def\ergs{{\rm erg}\,{\rm s^{-1}}}
\def\Zand{{Z~And}\,}

\usepackage[T1]{fontenc}
\usepackage{ae,aecompl}

\title[Disc instabilities in RS~Oph and Z~And]{Disc instabilities and nova eruptions in symbiotic systems: RS~Ophiuchi and Z~Andromedae}

\author[D. A. Bollimpalli, J.-M. Hameury, J.-P. Lasota]{D. A. Bollimpalli$^{1,4}$\thanks{E-mail: \href{mailto:deepika@camk.edu.pl}{deepika@camk.edu.pl}}
J.-M. Hameury$^{2,4}$\thanks{E-mail: \href{jean-marie.hameury@astro.unistra.fr}{jean-marie.hameury@astro.unistra.fr}}
J.-P. Lasota$^{3,1,4}$\thanks{E-mail: \href{mailto:deepika@camk.edu.pl}{lasota@iap.fr}}
\\
$^{1}$Nicolaus Copernicus Astronomical Center, Polish Academy of Sciences, ul. Bartycka 18, PL 00-716 Warsaw, Poland.\\
$^{2}$Universit\'e de Strasbourg, CNRS, Observatoire Astronomique de Strasbourg, UMR 7550, 67000 Strasbourg, France.\\
$^{3}$Institut d'Astrophysique de Paris, CNRS et Sorbonne Universit\'e, UMR 7095, 98bis Bd Arago, 75014 Paris, France.\\
$^{4}$Kavli Institute for Theoretical Physics, Kohn Hall, University of California, Santa Barbara, CA 93106, USA.\\
}

\date{Last updated; in original form}

\pubyear{2017}

\begin{document}
\label{firstpage}
\pagerange{\pageref{firstpage}--\pageref{lastpage}}
\maketitle

\begin{abstract}
Using the disc instability model for dwarf novae and soft X-ray transients, we investigate the stability of accretion discs in long-period binary systems. We simulate outbursts due to this thermal-viscous instability for two symbiotic systems, RS~Ophiuchi and Z~Andromedae. The outburst properties deduced from our simulations suggest that, although the recurrent nova events observed in RS~Oph are due to a thermonuclear runaway at the white dwarf surface, these runaways are triggered by accretion disc instabilities. In quiescence, the disc builds up its mass and it is only during the disc-instability outburst that mass is accreted on to the white dwarf at rates comparable to or larger than the mass-transfer rate. For a mass-transfer rate in the range $10^{-8}$ to $10^{-7}~\Msun$~yr$^{-1}$, the accretion rate and the mass accreted are sufficient to lead to a thermonuclear runaway during one of a series of a few dwarf nova outbursts, barely visible in the optical, but easily detectable in X-rays.
In the case of Z~And, persistent irradiation of the disc by the very hot white-dwarf surface strongly modifies the dwarf-nova outburst properties, making them significant only for very high mass-transfer rates, of the order of $10^{-6}~\Msun$~yr$^{-1}$, much higher than the expected secular mean in this system. It is thus likely that the so-called `combination nova' outburst observed in years 2000 to 2002 was triggered not by a dwarf-nova instability but by a mass-transfer enhancement from the giant companion, leading to an increase in nuclear burning at the accreting white-dwarf surface.
\end{abstract}




\section{Introduction}
Symbiotic stars are interacting binary systems in which a hot primary star accretes matter lost by an evolved-giant, secondary star. In most known cases the primary is a white dwarf, but several systems with a neutron-star primary have also been observed. These binary systems show several types of outburst. A small subset is observed as `slow novae' whose outbursts last for decades; another subset is formed by recurrent novae (RN) whose outbursts repeat on time-scales from several to several tens of years. Both types of nova outburst are known to result from thermonuclear runaways in the hydrogen-rich layer accumulated by accretion on the white dwarf surface. Table \ref{tab:table1} lists the properties of Galactic recurrent novae which have long orbital periods. The third type of outburst is known under the name of Classical Symbiotic Outburst (hereinafter CSO) and their origin is still a subject of debate \citep[see in e.g.][hereinafter S2006, and references therein]{sokoloski2006b}. 

\begin{table*}
\caption{Galactic recurrent novae with long binary orbital periods}
\centering
\begin{tabular*}{\textwidth }{@{\extracolsep{\fill}}l p{2cm} ccccc}
\hline
\hline
      object & 
      outburst year & 
        \multicolumn{1}{c}{$P_{\rm rec}$ /}  & 
        \multicolumn{1}{c}{$P_{\rm orb}$ /}  &
        \multicolumn{1}{c}{$M_{\rm WD}$ /}  &
        \multicolumn{1}{c}{$m_{\rm min}$ }  &
        \multicolumn{1}{c}{$m_{\rm max}$ }  \\
         &
         &
        \multicolumn{1}{c}{yr}  & 
        \multicolumn{1}{c}{d} &
        \multicolumn{1}{c}{M$_{\odot}$} &
        \multicolumn{1}{c}{(mag)} &
        \multicolumn{1}{c}{(mag)}\\
\hline

T CrB  &1866,1946 & $<$ 80 & 228 & $\gtrsim$ 1.35 & 2.5&9.8 \\
RS~Oph &1898,1907,1933,1945, & $<$ 22 & 453.6 &$\sim$  1.35 &4.8 & 11 \\
&1958,1967,1985,2006& & & & & \\
V745 Sco & 1937,1989,2014&  $<$ 52 & 510 & $\gtrsim$ 1.35 & 9.4&18.6  \\
V3890 Sgr &1962,1990 & $<$ 27 & 519.7 &$\gtrsim$ 1.35 &8.1 &15.5   \\
     \hline
\end{tabular*}
\begin{threeparttable}
\begin{tablenotes}
\item Here listed are the basic parameters of few galactic recurrent novae: The years outburst were observed and the recurrence period of the outburst are mentioned in the second and third columns. The next two columns shows the binary orbital period and mass of the white dwarf and the last two columns give the brightness in \textit{V} band measured during the outburst peak and quiescence.
References: \citep{Schaefer2010, Anupama2013}. 
 \end{tablenotes}
\end{threeparttable}

\label{tab:table1}
\end{table*}

In most cases the companion giant star underfills its Roche lobe and accretion on to the primary star occurs through stellar wind capture which does not guarantee the presence of an extended disc. Despite that, disc-involving, dwarf-nova type outbursts have been invoked as a possible source of, or a contributor to CSOs  \citep{duschl1986a, duschl1986b, mikolojaweska2002,sokoloski2006b} and even as an explanation of the outbursts of one of the recurrent novae, RS~Oph \citep[][]{KP09,Alex2011}.

Orbital periods of symbiotic stars range from years to decades which correspond to binary separations ranging from $10^{13}$ to $10^{15}$ cm, so that accretion discs (if present) in such wide binary systems are subject to the thermal-viscous instability even for huge mass-transfer rates, since the critical accretion rate, below which the disc is unstable, increases with a 2.4 to 2.6 power of the disc radius \citep[][]{Lasota2008}.

The typical size of the accretion disc in a dwarf nova ranges from about $10^{10}$ to 10$^{11}$ cm. Outbursts in such systems have been very well modelled by a disc instability model \citep[DIM; see][for a review of the model]{Lasota2001}. There have not been many attempts to apply the DIM to the description of the large discs of symbiotic stars (SBs), with the notable exceptions of those by \citet{duschl1986a, duschl1986b} and \citet{Alex2011}. In two papers, \citet{duschl1986a, duschl1986b} studied the stationary and time-dependent behaviour of accretion discs around a main-sequence star and suggested that the light curves of SBs can be explained by disc instabilities. Since the accreting primary was a main-sequence star and the transition fronts were approximated as quasi-stationary, these models are not directly applicable to the study of outbursts in SBs as we know them at present \citep{sokoloski2017}. The simplified  code used by  \citet{Alex2011} in their study of RS~Oph outbursts requires serious improvements in order to be able to describe the physics of accretion disc outbursts (Graham Wynn, private communication; see also Sect. 3.1.1). The DIM code developed by \citet{HMDLH98} and \citet{dubus1999} has been used to describe outbursts of irradiated low-mass X-ray binary (LMXB) discs extending up to radii of a few $10^{11}$cm \citep{DubusPhD,dubus2001} but larger configurations have not been yet explored.

In this article we use our DIM code to test two hypotheses involving disc instabilities in the context of SBs. First, inspired by \citet{KP09} and \citet{Alex2011}, we examine the role of dwarf-nova type of instabilities in the 2006  outburst of the recurrent nova RS~Oph. In particular we test the hypothesis \citep[][Cannizzo, private communication]{Alex2011} that the thermonuclear runaway in this and other similar systems could be triggered by a single dwarf nova instability. Secondly, we address the `combination nova' scenario (S2006) according to which a CSO observed in the prototypical SB Z~And is due to an increase of thermonuclear shell burning triggered by accretion disc instability. We also study the possibility of `pure' dwarf-nova outbursts in shell-burning SBs. A direct comparison of the light curves generated by the DIM code with the observational data is not appropriate because this would require to couple the DIM code with a code describing the thermonuclear runaway at the white dwarf surface. Instead, for each model we consider, we compare the model predictions of the general outburst characteristics (outburst duration and amplitude, recurrence time) with the observed values.

\section{The model}

We use the version of the DIM described by \citet{HMDLH98} with added heating due to stream impact and the dissipation by the tidal torques as described by \citet{BM01}. The code we have developed solves the standard disc equations in the radial dimension \citep[mass, angular momentum and energy conservation, see e.g.][]{Lasota2001} of a geometrically thin, optically thick disc in order to follow its thermal and viscous evolution. Because the disc is assumed to be geometrically thin, the radial and vertical dimensions decouple, and the viscous heating as well as the radiative cooling terms can be calculated as a function of local quantities (surface density, central disc temperature, radius) only. The code is implicit and uses an adaptive grid so that the heating and cooling fronts are well resolved; the inner and outer edges of the disc are allowed to vary in time (see below). Time derivatives are calculated using a Lagrangian scheme. The non-linear equations are solved using a generalized Newton--Raphson method.

The heating and cooling terms are calculated by solving the vertical structure of the accretion disc, assuming hydrostatic and thermal equilibrium. The equations are very similar to those used for stellar structure. The main difference is the heating term, due to viscous instead of nuclear processes. This enables to determine the temperature profile, and in particular the central and effective temperature as a function of the local surface density $\Sigma$ and radius, for a given viscosity parameter $\alpha$. The disc is not necessarily in thermal equilibrium though; as noted by \citet{HMDLH98}, this introduces an additional term in the heat transport equation, which is supposed to have the same behaviour as the viscous heating term, i.e. is proportional to pressure. The disc vertical structure can then be determined for any position and $\Sigma$ when the disc is not in thermal equilibrium, but $\alpha$ is no longer given; the central temperature then becomes a free parameter. In order to speed-up the resolution of the radial time-dependent equations, effective temperatures are pre-calculated on a grid in radius, $\Sigma$ and central temperature and trilinear interpolation is performed on this grid.

Throughout this paper, unless otherwise noted (model RSB-A, see sect. 3), we take for the Shakura--Sunyaev viscosity parameters $\alpha_{\rm c} = 0.02$ for a cold neutral disc and $\alpha_{\rm h} = 0.1$ when the disc is hot and ionized.

\subsection{Disc irradiation}
\label{sec:irrad}

For irradiated discs we use the version of the code (iDIM) described by \citet{dubus1999,dubus2001,Lasota2008} in the context of low-mass X-ray binaries and by \citet{HLD99} in the slightly different context of cataclysmic variables. Irradiation does not affect the time-dependent radial equations, but only modifies the vertical structure of the disc by changing the relation $F_z = \sigma T_{\rm eff}^4$ into $F_z + F_{\rm irr}= \sigma T_{\rm eff}^4$.

The flux $F_{\rm irr}$ irradiating the outer disc regions is written as
\begin{equation}
\label{eq:flirrad}
F_{\rm irr}\equiv \sigma_{\rm SB}T_{\rm irr}^4 = {\cal C}\frac{L}{4\pi R^2},
\end{equation}
where $R$ is the radial coordinate, $\sigma_{\rm SB}$ the Stefan-Boltzmann constant, and ${\cal C} <1$ is a constant encapsulating the irradiation geometry, albedo etc. This very simple description of disc irradiation gives satisfactory results when describing outer-disc irradiation in X-ray binaries, in particular in transient systems \citep{dubus2001, tetarenko2018a}. In this case $L=L_{\rm acc}=\eta \dot M c^2$, where $\eta$ is the accretion radiation-efficiency. The ratio of irradiation to accretion (`viscous') flux is, in the steady state case
\begin{equation}
\label{eq:flratio}
\frac{F_{\rm irr}}{F_{\rm acc}}=\frac{4}{3}{\cal C}\eta \frac{R}{R_{\rm S}F},
\end{equation}
where $R_{\rm S}=2GM_1/c^2$ is the Schwarzschild radius, $M_1$ being the primary mass, and
\begin{equation}
\label{eq:flacc}
F_{\rm acc}\equiv \sigma_{\rm SB}T_{\rm eff}^4 = \frac{3GM_1\dot M}{8\pi R^3}.
\end{equation}
For ${\cal C} \approx 10^{-3}$ and $\eta \approx 0.1$, used in models of X-ray transients \citep{dubus2001}, $F_{\rm irr}/{F_{\rm acc}}\approx 10^{-4} R/R_{\rm S}$ and $T_{\rm irr} > T_{\rm eff}$ for $R > 10^4 R_{\rm S}$. However, for white dwarfs $\eta \approx 10^{-5}$ to $10^{-3}$, and even for very large discs, self-irradiation of the outer disc is never in practice important because it would require disc radii $R > 10^6$ to $10^8 R_{\rm S} \approx 4 \times 10^{11}$ to $10^{13}$ cm. Only for massive white dwarfs and ${\cal C} \approx 10^{-2}$ self-irradiation of its outer region could play a role in the disc structure and evolution (see below the discussion in the case of RS~Oph).

However, in some systems such as symbiotic stars and supersoft X-ray sources, steady thermonuclear burning of matter accumulated at the white dwarf surface can be important and even dominate accretion disc irradiation.
In this case 
\begin{gather}
\frac{F_{\rm irr,*}}{F_{\rm acc}} \approx \frac{2}{3} \frac{{\cal C}L_*}{GM_1\dot M/R_*} \frac{R}{R_*} \nonumber \\
\approx 0.5 \left(\frac{\Msun}{M_1}\right)\left(\frac{R} {10^8\rm cm}\right)
\left(\frac{{\cal C}L_*}{10^{37} \rm erg\, s^{-1}}\right)\left( \frac{10^{19}\rm g\,s^{-1}}{\dot M}\right),
\label{eq:flrationuc}
\end{gather}
where $R_*$ is the white dwarf radius. In the symbiotic star Z~And, for example, the steady luminosity of thermonuclear burning is about $10^3\,\Lsun$. So even for very high accretion rates the outer disc surface temperature is dominated by irradiation from the central source. Since constant irradiation is independent of the accretion rate, it impacts the whole disc outburst cycle because, contrary to the accretion-self-irradiation case, it is also present during quiescence. Notice that in Eq.~(\ref{eq:flrationuc}), contrary to Eq.~(\ref{eq:flratio}), the ratio $F_{\rm irr,*}/F_{\rm acc}$ depends on the accretion rate and \textsl{decreases} with increasing $\dot M$.

One should note, however, that it is not clear in what conditions Eq.~(\ref{eq:flirrad}) applies to discs around white dwarfs, in particular what value of  ${\cal C}$ should be used. In X-ray binaries, for which there is direct evidence of outer disc irradiation, a `reasonable' value for ${\cal C}$ is about a few $10^{-3}$ \citep[][see, however, \cite{tetarenko2018b}]{dubus1999,coriat2012}. Note that ${\cal C}$ used here differs from the ${\cal C}$ as defined by \citet{dubus1999} in that it does not include the efficiency parameter $\eta $.

Eq.~(\ref{eq:flirrad}) assumes that the disc is irradiated by a point source. This assumption is no longer true close to the white dwarf and irradiation has to be described by \citep[see e.g.,][]{HLD99}
\begin{equation}
\label{eq:fliradwd}
F_{\rm irr,S}\equiv \sigma_{\rm SB}T_{\rm irr,*}^4 = (1- \beta)\left[\sin^{-1}\rho - \rho\sqrt{1 - \rho^2}\right]\frac{T^4_{*}}{\pi},
\end{equation}
where the subscript $*$ refers to quantities measured at the white dwarf surface; $\rho = R/R_*$.

At large distances, $F_{\rm irr,S}$ as given by Eq.~(\ref{eq:fliradwd}) varies as $R^{-3}$ and is much smaller than what one would get using Eq.~(\ref{eq:flirrad}); conversely, close to the white dwarf, Eq.~(\ref{eq:flirrad}) underestimates the irradiation flux.

In what follows, when the ratio of the disc scale-height to stellar radius $H/R_* >1$, we use Eq.~(\ref{eq:flirrad})  to describe irradiation, whereas in the opposing case, $H/R_* <1$, Eq.~(\ref{eq:fliradwd}) is used.

\subsection{Boundary conditions}

Since radiation pressure can be dominant during outbursts, and gravity could potentially vary within the extended disc photosphere, the usual photospheric boundary condition $\kappa P_{\mathrm g}=2/3g_{z}$ is replaced by \citet{HL2009} with
\begin{equation}
\kappa \left(P_{\mathrm g}+ \dfrac{1}{2} P_{\mathrm{rad}}\right) = \dfrac{2}{3}g_{z}\left( 1+ \dfrac{1}{\kappa \rho z}\right),
\label{eq:brp}
\end{equation}
where $\kappa$ is the Rosseland mean opacity, $g_{z}$ is the vertical component of gravity, $\rho$ the density, $P_{\mathrm g}$ and $P_{\mathrm{rad}}$ are the gas and radiation pressure respectively.

In all models the disc is truncated at a magnetospheric radius given by
\begin{equation}
R_{\rm mag} = 5.13 \times 10^{8} \mu_{30}^{4/7} \left( \frac{M_1}{\Msun} \right)^{-1/7} \left( \frac{\dot{M}_{\rm acc}}{10^{16} \; \rm g\;s^{-1}} \right)^{-2/7} \; \rm cm,
\label{eq:rmag}
\end{equation}
where $M_1$ is the accretor mass in solar units, $\dot{M}_{\rm acc}$ is the accretion rate on to the white dwarf. This differs from $\dot{M}_{\rm tr}$ if the system is not steady, and $\mu_{30}$ is the white dwarf magnetic moment in units of $10^{30}$ G cm$^3$ \citep{fkr02}. 

\begin{figure}
\includegraphics[scale=0.42]{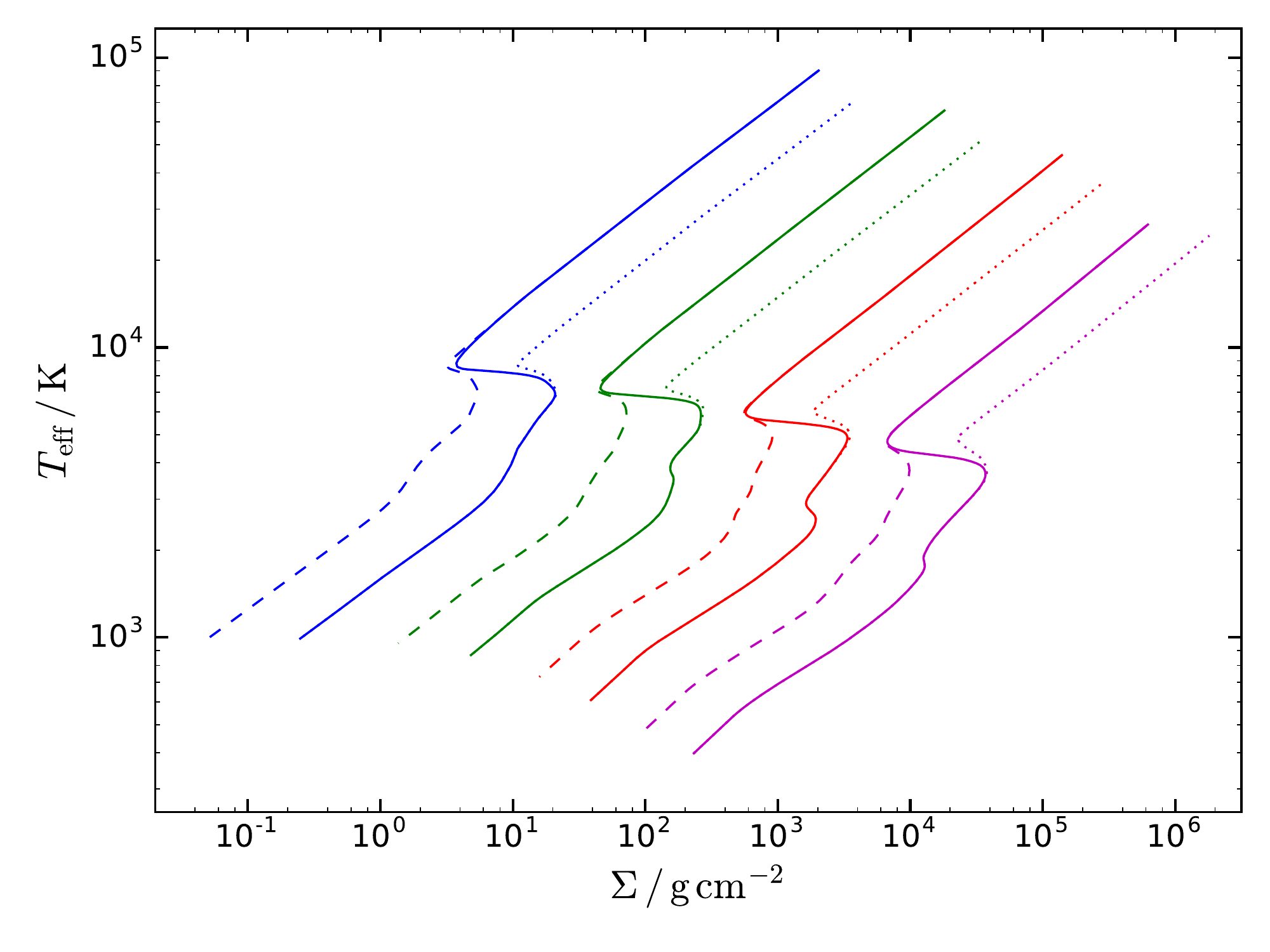}
\caption{ $\Sigma - T_{\mathrm{eff}}$ S curves computed for a 1.35 $\Msun$ WD are plotted for different radii. The set of four curves, from left to right, in each linestyle represent the S curves at radii $r = 10^{9},~10^{10},~10^{11}$ and $10^{12}$ cm respectively. Solid lines represent the S curves attained from temperature dependent $\alpha$. Dashed and dotted lines represent S curves for fixed $\alpha = 0.1$ and $\alpha = 0.01$ respectively (we use these values for direct comparison with Fig. 1 of \citet{Alex2011}; see Sect. \ref{rsoph}.)}
\label{fig:fig1}
\end{figure}

\subsection{S--curves}

On the $\Sigma$--$T_{\rm eff}$ plane, accretion disc steady-states, at a given radius, form celebrated `S--curves'. The upper and lower branches are stable; the intermediate branch is thermally and viscously unstable. As mentioned earlier, these S--curves are calculated by solving the disc vertical structure. The bends in the S--curves define the critical values of the parameters at which the stability properties of a local disc ring are changing. The numerical fits to these values can be found in Appendix \ref{sect:critvalues}.

Fig. \ref{fig:fig1} shows examples of such S--curves we obtain for various radii. Fits to the critical surface densities, effective temperatures and corresponding accretion rates are given in Appendix A. These fits are valid for radii in the range 10$^8$ to 10$^{12}$ cm and for $\alpha$ in the range 10$^{-4}$ to 1. As can be seen, the S--curves do not look very different from those found for smaller discs; the extension to large radii does not change the overall shape of the S--curves, but merely increases typical surface densities and slightly decreases the corresponding effective temperatures. They do, however, differ quite significantly from those used by \citet{Alex2011}, see their fig. 1. We return to this point in Sect. 3.1.1. Note also that the extent of the unstable branch is largely dominated by the jump in the viscosity parameter \citep{HMDLH98}.

\subsection{Light-curves}

When calculating the light-curves we include the contributions from the thermal emission of the white dwarf, the (irradiated) secondary star and the hot spot where the mass-transfer stream hits the disc \citep[see][]{shl2003}. For the shell-burning sources we model their emission as that of a black-body with the shell luminosity $L_{\rm nuc}$. The contribution of the red giant is important and more difficult to evaluate, particularly because it may or may not fill its  Roche lobe. Specific details for RS~Oph and Z~And are given in the corresponding sections.

We also account for interstellar reddening that decreases the observed optical fluxes; the visual extinction is $A_{\rm V}=3.1E(B-V)$, where $E(B-V)$ is the colour excess. For RS~Oph, we take $E(B-V) = 0.73$ \citep{snijders1987}. For Z~And, we consider the minimum $E(B-V)=0.27$ as did S2006; values for $E(B-V)$ found in the literature are in the range of 0.27 to 0.35.

\begin{figure*}
\begin{center}
\includegraphics[width=\textwidth]{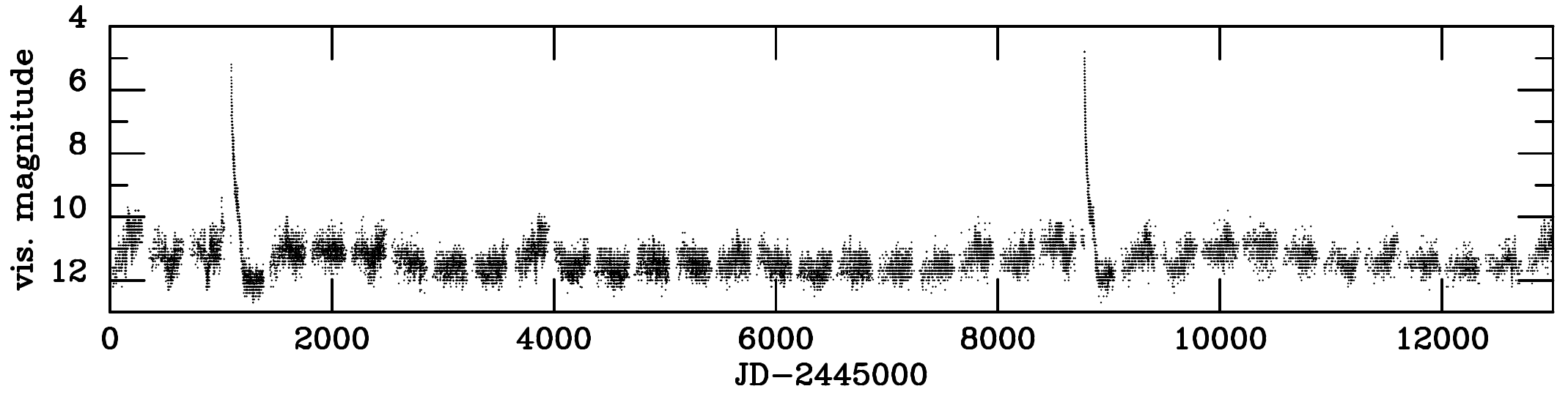}
\caption{RS~Oph visual light curve (data from AAVSO)}
\label{fig:lc_aavso_rsoph}
\end{center}
\end{figure*}

\section{RS~Ophiuchi}
\label{rsoph}

RS~Oph is a symbiotic RN with a recurrence time of about 20 yr. The binary has an orbital period of 453.6 days, the mass ratio $q\equiv M_2/M_1=0.6$, and the eccentricity $e=0$ \citep{brandi09}. The low-mass ejecta of the 2006 outburst \citep[approximately in the range $10^{-7}$ to $10^{-6}$ $\Msun$;][]{sokoloski2006a} and their high velocity \citep[4000 km~s$^{-1}$;][]{Das2006,Anupama2006} suggest a white dwarf mass close to the Chandrasekhar limit \citep[see e.g.][for a study of the dependence of the ejecta characteristics upon the system parameters]{Yaron2005}. Other arguments also point towards a high-mass primary \citep{MS2017}. Therefore the mass of the K7 giant \citep{MuersetSchmid1999} companion can be estimated to be $M_2\approx 0.8\,\Msun$, the inclination is about $50^o$, and orbital separation $a\approx 2.2\times 10^{13}$cm.

Fig. \ref{fig:lc_aavso_rsoph} shows the long term light curve of RS~Oph in visual over the period 1982 to 2017, obtained using AAVSO data. The visual magnitude varies from 11--12 in quiescence to 4.8 at maximum. During weeks 6 to 10 of the last, 2006 outburst, RS~Oph was a supersoft X-ray source with a luminosity in the 0.2 to 1.0 keV range close to the Eddington luminosity and a temperature of the white dwarf photosphere close to $8\times 10^5$K \citep{Nelsonetal2008}. Nuclear burning ended around day 69 after outburst. The nova shell ejected by RS~Oph was asymmetrical and contained a jet-like structure. Although this feature of the outburst motivated \citet{KP09} to assert that it was not thermonuclear but resulting from a dwarf-nova instability, the clear signature of nuclear burning observed in RS~Oph during the supersoft state makes such a possibility highly improbable.

The giant companion star is probably close to filling its Roche-lobe. Estimates of the mass-loss rate vary from about $10^{-8}\Msun$~yr$^{-1}$ \citep{Iijima2008} to $10^{-6}\Msun$~yr$^{-1}$ \citep{Evansetal2007}. Recent estimates by \citet{Bootheal2016} give $5\times 10^{-7}$ $\Msun$~yr$^{-1}$.

Until \textsl{GAIA} determined the parallax of RS~Oph that corresponds to a distance of $2.26 \pm .27$ kpc \citep{gaiaDR2}, the distance of 1.6 kpc to RS~Oph based on \citet{Bode87} had usually been assumed but distances as low as 540 pc \citep{Monnier06} and as high as 2.45 kpc \citep{Rupenetal08} had been claimed. For comparison with previous studies, we use here the value of 1.6 kpc, and discuss at the end of this section the impact of the updated distance. The quiescent accretion rate is variable. At day 241 after the 2006 outburst, when optical flickering had resumed indicating re-formation of an accretion disc, the accretion rate was estimated to be within $10^{-10}$ and $10^{-9}\,\Msun$~yr$^{-1}$, depending on the assumed model of mass-loss rate from the secondary \citep{Wortersetal07}. However, on days 537 and 744, X-ray observations suggested  accretion rates between $ 7\times 10^{-11}$ and $5\times 10^{-9}\,\Msun$~yr$^{-1}$, depending on the emission model and assuming a 1.6 kpc distance \citep{Nelsonetal2011}. At the true GAIA distance of 2.26 kpc the accretion rates would be respectively $ 1.4\times 10^{-10}$ and $1.0\times 10^{-8}\,\Msun$~yr$^{-1}$. The presence of an absorbed optically thick boundary layer would allow accretion rates up to $3.2\times 10^{-8}\,\Msun$~yr$^{-1}$ \citep[for a 1.6 kpc distance;][]{Nelsonetal2011}. However, observations by \textsl{ROSAT PSPC} of RS~Oph six and seven years after the 1985 outburst implied quiescent accretion rates in the range $10^{-12}$ to few $10^{-11}\,\Msun$~yr$^{-1}$ \citep{Orio1993,Orioetal2001}. Similar results were obtained by \citet{Mukai2008} in 1997. Such low accretion rates have been considered as `completely at odds with the short nova outburst recurrence time ($\sim $20 years)' \citep{Nelsonetal2011}.

The reason is that a thermonuclear runaway recurrence time of 20 yr, for a white dwarf close to the Chandrasekhar limit requires an average accretion rate of about $10^{-8}\Msun$~yr$^{-1}$ \citep{Yaron2005}. Therefore the highest estimates of both the companion mass-loss rate and the accretion rate on the white dwarf are consistent with a nova outburst occurring every twenty years. However, since these estimates are model and distance dependent, the consistency between the observed properties of RS~Oph outbursts and the standard nova-outburst model cannot be assumed with certainty. Since the accretion disc, the presence of which seems to be confirmed by the observed optical flickering and by the necessity of a high accretion efficiency, may extend to a distance large enough to make it thermally and viscously unstable, it is worth studying if this instability gives rise to large scale outbursts and if these outbursts, if occurring, can drop enough mass on to the white dwarf to trigger a thermonuclear runaway while having, according to the DIM, very low quiescent accretion rates. We also wish to test the claim by \citet{Alex2011} that RS~Oph outbursts have their origin not in thermonuclear runaways, but are due to stable hydrogen burning  at the white dwarf surface, triggered by dwarf-nova outbursts recurring every 20 years.

\subsection{Dwarf-nova outbursts of RS~Oph}
\label{subsect: dnrsoph}

\begin{table}
\caption{Models of RS~Oph dwarf-nova outbursts}
\centering
\begin{tabular}{lccc}
\hline
\hline
\vspace{2mm}
 Model & $R_{\mathrm{in}}(qsc) /$  & $R_{\mathrm{D}}$ /&  $\dot{M}_{\rm{tr}}$ / \\
       & $10^{10}$ cm              &  $10^{10}$ cm     & $\Msuny$ \\
\hline
RSB-6 & $0.215$&$256.25$& $10^{-6}$\\
RSB-7 & $0.215$&$256.25$&$10^{-7}$\\
RSB-8 &$0.215$ &$256.25$&$10^{-8}$ \\
RSB-A$^*$&$1.0$ &$253.56$&$10^{-6}$ \\
\hline
\end{tabular}
\begin{threeparttable}
\begin{tablenotes}
\item
$R_{\rm in}(qsc)$ is the inner disc radius during quiescence, $\dot{M}_{\rm tr}$ is the mass-transfer rate from the secondary and $R_{\rm D}$ is the outer disc radius.   
$^*$ -- model closest to \citet{Alex2011}. 
 \end{tablenotes}
\end{threeparttable}
\label{tab:table_models_RS}
\end{table}

\begin{table*}
\caption{Disc outbursts properties for RS~Oph models.}
\begin{threeparttable}
\centering
\begin{tabular*}{2 \columnwidth }{@{\extracolsep{\fill}} l c c c c c c c c}
\hline
\hline \\ [-1.8ex]
      Model & 
      \multicolumn{1}{c}{$\dot{M}_{\rm in}$ (qsc)/}&
      \multicolumn{1}{c}{$\dot{M}_{\rm in}$ (peak)/} & 
      \multicolumn{1}{c}{$m_{\rm V}$}  &
      \multicolumn{1}{c}{$m_{\rm V}$}  &
      \multicolumn{1}{c}{$\Delta M$ /}  &
      \multicolumn{1}{c}{$\tau_{\mathrm{rec}}$ /}  &
      \multicolumn{1}{c}{ $\tau_{\mathrm{ob}}$ /} &
      \multicolumn{1}{c}{ $R_{\mathrm{f,max}}$ /} \\

         &
        \multicolumn{1}{c}{$\Msun$yr$^{-1}$} &
        \multicolumn{1}{c}{ $\Msun$yr$^{-1}$} &
        \multicolumn{1}{c}{(qsc)}  & 
        \multicolumn{1}{c}{(peak)} & 
        \multicolumn{1}{c}{ $\Msun$} & 
        \multicolumn{1}{c}{ d} & 
        \multicolumn{1}{c}{ d}&
        \multicolumn{1}{c}{ $10^{10}$ cm} \\ [0.8ex]
   
\hline \\ [-1.8ex]
RSB-6&$3.8 \times 10^{-12}$&$4.15 \times 10^{-6}$  &11.03 &9.08 &$10^{-6}$ &556 & 258&56 \\
RSB-7&$3.8 \times 10^{-12}$&$6.3 \times 10^{-7}$ & 11.03 & 10.48 & 6.8$\times 10^{-8}$& 335 & 115&27\\
RSB-8&$ 3.8\times 10^{-12}$&  $9.3 \times 10^{-8}$ &11.03 & 10.9 & $4.5 \times 10^{-9}$&205 &50 &12.7 \\
\hline \\ [-1.8ex]
RSB-A& $4.5 \times 10^{-12}$&$9.6 \times 10^{-6}$  &11.03  &8.25 &3.35 $\times 10^{-6}$ & 1190 & 260&70 \\
\hline
\end{tabular*}
\begin{tablenotes}
\item The first column lists the model reference, the second and third ones give the minimum quiescent and outburst peak of the mass accretion rate. The fourth and fifth columns give the apparent visual magnitude during quiescence and outburst peak respectively. The sixth column gives an estimate of the amount of mass accreted on to the white dwarf during one outburst. The next two columns denote the recurrence time for the outburst and the outburst duration respectively. The last column gives  the largest distance reached by the heating fronts.
\end{tablenotes}
\end{threeparttable}
\label{tab:table_RSDIM}
\end{table*}

It is not always realized that the presence of a thermally and viscous unstable disc is only a necessary condition for the occurrence of dwarf-nova type outbursts. For example, accretion discs in AGNs, even when unstable, do not show large scale outbursts for reasons explained by \cite{HL2009}. It is therefore not obvious a priori what is the response of a very large disc to a thermal-viscous instability. To study this problem we calculated a series of models for which parameters are given in \autoref{tab:table_models_RS}.

\subsubsection{Unirradiated discs}

First, we consider three models for a system with parameters appropriate for RS~Oph (RSB-6, RSB-7, RSB-8) corresponding to three different mass-transfer rates from the secondary of $10^{-6}$, $10^{-7}$ and $10^{-8}~\Msun\,{\rm yr}^{-1}$. To these three models we add RSB-A which is the closest we could get to the model considered by \citet{Alex2011}. 

In all models the mass of the white dwarf is $1.35 \Msun$ \citep[see e.g.][]{HK07} and the assumed distance is 1.6 kpc.
To be compatible with \citet{Alex2011}, we assume the secondary's temperature to be $3500$ K \citep{schaefer2009}. In order to have a truncated disc, we assume arbitrarily a white dwarf magnetic moment of $2 \times 10^{30}$ G ${\rm cm}^3$. Truncation of the inner disc is necessary to avoid multiple small outbursts which are inevitable when the inner disc radius is as small as that of a 1.35 $\Msun$ white--dwarf radius and are just a numerical nuisance.

Since the giant companion to RS~Oph is not supposed to fill its Roche lobe, it is not clear what the structure of the accretion flow in this system is. In particular what is the size of the accretion disc that apparently is present during quiescence. For the assumed binary parameters the circularization radius for the Roche-lobe filling case is $R^{\rm RL}_{\rm circ}= 2.37 \times 10^{12}$cm \citep{Lubow75}, while in the case of wind accretion it is $R^{\rm W}_{\rm circ}< 2.2 \times 10^{11}\lambda(a)^{-4}$cm, where $\lambda \approx 1$ \citep{fkr02}.  In our code the average outer disc radius is determined by the tidal torque
\begin{equation}
{\cal T}_{\mathrm{tid}} = c_{\mathrm{tid}} \Omega_{\mathrm{orb}} r \nu \Sigma \left(\dfrac{r}{a}\right)^5,
\label{tid}
\end{equation}
where $\nu$ is the kinematic viscosity coefficient and $c_{\mathrm{tid}}$ is a numerical coefficient which modulates the tidal truncation radius and provides an average value of the disc outer radius, $R_{\mathrm{D}}$ when the disc is steady. Here we use $c_{\mathrm{tid}}=3.0 \times 10^{9}$ such that $R_{\mathrm{D}}$ is of the order of $10^{12}$ cm, which is appropriate for both cases of mass-transfer from the giant companion. 

We also simulate the RS~Oph system using the same parameters as \citet{Alex2011}, referred to as model RSB-A in \autoref{tab:table_models_RS}. The masses of the white dwarf and of the secondary are identical to those in RSB-6, 7 and 8, but, in order to have the same circularization radius, $R_c \approx 1.74\times 10^{12}$ cm and disc size (slightly larger than $R_c$) as theirs, we assume the orbital period to be 287.5 days. As in \citet{Alex2011}, the disc is truncated at a fixed inner radius of $10^{10}$ cm. We also use $\alpha_{\rm c} = 0.01$ as in their model.
 
\begin{figure}
\begin{center}
\includegraphics[width=\columnwidth]{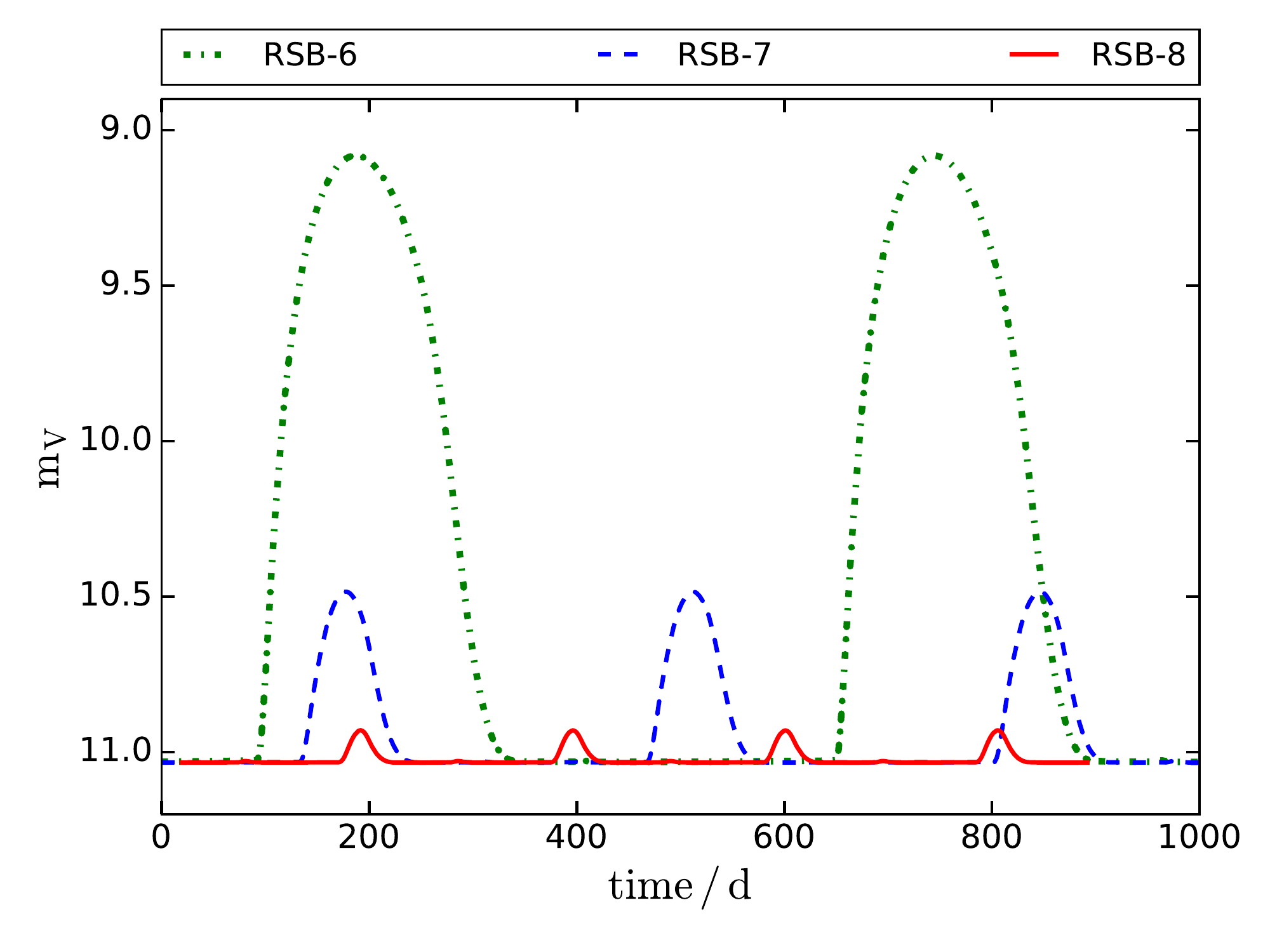}
\caption{Optical light curves for the models RSB-6 (dash-dotted green curve), RSB-7 (dashed blue curve), RSB-8 (solid red curve).}
\label{fig:lc_rsoph}
\end{center}
\end{figure}

\begin{figure}
\includegraphics[width=1\columnwidth]{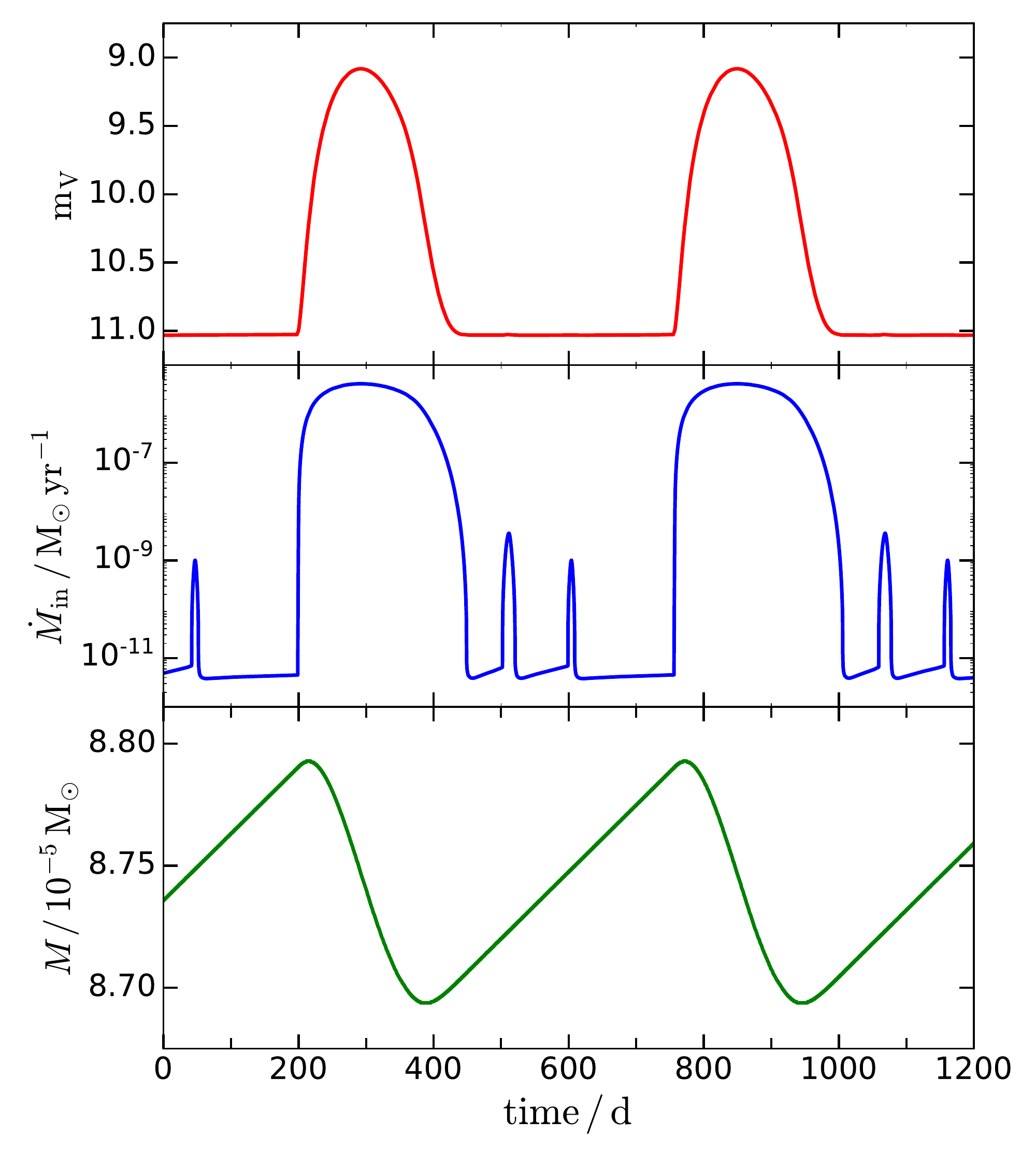}
\caption{From top to bottom: visual magnitude, accretion rate on to the white dwarf and disc mass for model RSB-6.}
\label{rsoph_mdot}
\end{figure}

Fig. \ref{fig:lc_rsoph} shows the light curves obtained for the three models considered here, and 
\autoref{tab:table_RSDIM} provides the main outburst characteristics for the RS~Oph dwarf-nova outbursts. As the mass-transfer rate increases, the recurrence time between outbursts increases and the outbursts last longer too. In none of the models do heating fronts reach the disc outer edge, and the outer radius of the disc does not vary during the outbursts contrary to what is seen in smaller discs. During quiescence the typical accretion rate on to the white dwarf $\dot{M}_{\rm in } \gtrsim 10^{-12} ~\Msun~{\rm yr}^{-1}$. The larger the mass-transfer rates, the larger the peak value of $\dot{M}_{\rm in }$ of an outburst. Fig. \ref{rsoph_mdot} shows the time variations of the visual magnitude, the mass accretion rate (a proxy for the X-ray emission), and the total disc mass for model RSB-6, which has the longest recurrence time. The small intermediate outbursts are not detectable in visible light, but could be marginally detectable in X-rays.
 
Because the peaks of $\dot{M}_{\rm in }$, listed in \autoref{tab:table_RSDIM}, are more than three orders of magnitude larger than in quiescence, most of the mass is accreted on to the white dwarf during outbursts only. \autoref{tab:table_RSDIM} also shows that the visual amplitude of the outbursts is much smaller than the amplitude observed in RS~Oph: less than 2  vs $6.5$ mag. Also, the recurrence times of the outbursts for models RSB-6, RSB-7, RSB-8 (from 0.5 to 1.5 yr) are much shorter than the observed 20 yr. Therefore, contrary to the suggestion of \citet{KP09}, the DIM as such cannot account for the recurrent outbursts of RS~Oph.

Finally, we comment on model RSB-A. Our results differ widely from those of \citet{Alex2011}: the recurrence time we obtain is shorter than what they get for a mass-transfer rate of $10^{-6}~\Msun$~yr$^{-1}$ by a factor four and their outburst duration is longer; as a consequence, in their case the mass accreted on to the white dwarf in a single outburst is an order of magnitude larger than the value $3 \times 10^{-6}~\Msun$ we obtain for model RSB-A. These differences may be attributed to the way the disc's vertical structure is treated. \citet{Alex2011} assume that the disc is vertically isothermal, and the cooling rates are computed using a one-zone approximation. This approximation is no longer valid if the opacities vary significantly with height in the disc, which is the case during outbursts. Moreover, their effective temperatures are computed with a relation which only holds for large optical depths (they use $T_{\rm eff} = T_{\rm c} \tau^{-1/4}$), which is inconsistent with their estimate for the cooling term. In our models, we fully solve the vertical structure. Comparing our Fig. \ref{fig:fig1} where we plot the S--curves for RS~Oph computed using the same parameters as \citet{Alex2011}, with their Fig. 1 illustrates the substantial differences between the two approaches. Among other, the \citet{Alex2011} effective temperatures at the critical surface densities are lower than ours by factors of 3 to 4. This significantly alters the results. One should also note that truncating the disc at 10$^{10}$ cm is rather excessive. Magnetic truncation would imply a magnetic moment larger than $3 \times 10^{33}$ Gcm$^3$, i.e., large enough for pulsations at the spin period to be observed. 

\subsection{`Combination recurrent-nova' outbursts of RS~Oph}

One should still examine the possibility that, by dropping enough mass on the white-dwarf surface, dwarf-nova outburst(s) trigger a nova outburst. To start a thermonuclear runaway (TNR) in a system with RS~Oph parameters, one has to accumulate on the white dwarf surface a mass of $M_{\rm ign}\approx 4\times 10^{-7}\,\Msun$ \citep[][]{Wolf2013}.

We have considered this possibility for our three typical models.
\begin{description}
\item{-- RSB-6:} In this model the mass accreted by the white dwarf during a single outburst $\Delta M \approx 10^{-6}\,\Msun$ is larger than $M_{\rm ign}$, so that one dwarf-nova outburst could, in principle, trigger a nova event.
The mass $M_{\rm ign}$ is accumulated during the first 90 days of the disc outburst but, by the time this critical mass has been accreted, $\dot{M}_{\rm in} $ exceeds the stable nuclear burning limit, which is $\dot{M}_{\rm stable} = 6 \times 10^{-8} \Msun$~yr$^{-1}$ \citep{Wolf2013}. $\dot{M}_{\rm in} $ also exceeds  the limit $\dot{M}_{\rm RG} = 3 \dot{M}_{\rm stable}$ above which hydrogen can be burned as quickly as it is accreted, so that the hydrogen mass increases; therefore a thermonuclear runaway can occur only during the decay from the outburst maximum, as soon as the accretion rate falls below $\dot{M}_{\rm stable}$, which happens at day 140 after the outburst has been triggered. The early phases of the outburst, corresponding to a change in visual magnitude of about 2 could be just visible in the optical \citep{adam11} but would be easy to detect in X-rays since the accretion rate increases by six orders of magnitude. Of course, for such a scenario to work, the rate at which mass is transferred to the disc must be of order of $10^{-6}\Msun$~yr$^{-1}$ which implies that most of the mass lost by the companion must end up in the accretion disc. This is far for being guaranteed but this tight requirement could be loosened if disc self-irradiation were taken into account allowing more mass accreted with lower accretion rate (see below). The quiescent accretion rate corresponds to 1991, 1992 \textsl{ROSAT PSPC} and 1997 \textsl{ROSAT} post-1985 outburst observations. The higher post-2006 accretion rates observed by \textsl{Chandra} and \textsl{XMM-Newton} could coincide with `mini-outbursts' observed in the model light-curve. The main problem with this scenario is that, because each dwarf nova outburst is able to trigger a TNR, the recurrence time is very short, of order of 1.5~yr, more than an order of magnitude shorter than observed.

\item{-- RSB-7:} For this model  $M_{\rm ign}$ is accumulated during six dwarf-nova outbursts in about five years. The quiescent accretion rate is the same as in RSB-6, because of the same disc truncation. Between thermonuclear eruptions, the accretion rate would increase due to dwarf-nova outbursts that, in this case, are unable to trigger TNRs. They would be invisible in optical but could be detected in X-rays.

\item{-- RSB-8:} For this model $M_{\rm ign}$ is accumulated during 80 dwarf-nova outbursts during a period of around 45 yr. As for the preceding model, these disc outbursts would be unobservable in optical but the increased accretion rate could be detected through X-ray observations during most of the 50 days of the activity.
\end{description}

We conclude that a model with an accretion rate between those of RSB-7 and RSB-8 (mass-loss rate from the companion of a few times $10^{-8}\,\Msun$~yr$^{-1}$) can explain most of the properties of the RS~Oph recurrent nova outbursts. The mass necessary for a TNR is accumulated during 20 yr in a series of 15 to 20 dwarf nova outbursts. Such a scenario has the advantage of naturally explaining the discrepant X-ray observations at different quiescent epochs of the RS~Oph outburst cycle: low values correspond to the quiescent disc phase, while high X-ray luminosities to one of the dwarf nova outbursts.

\begin{figure}
\includegraphics[width=\columnwidth]{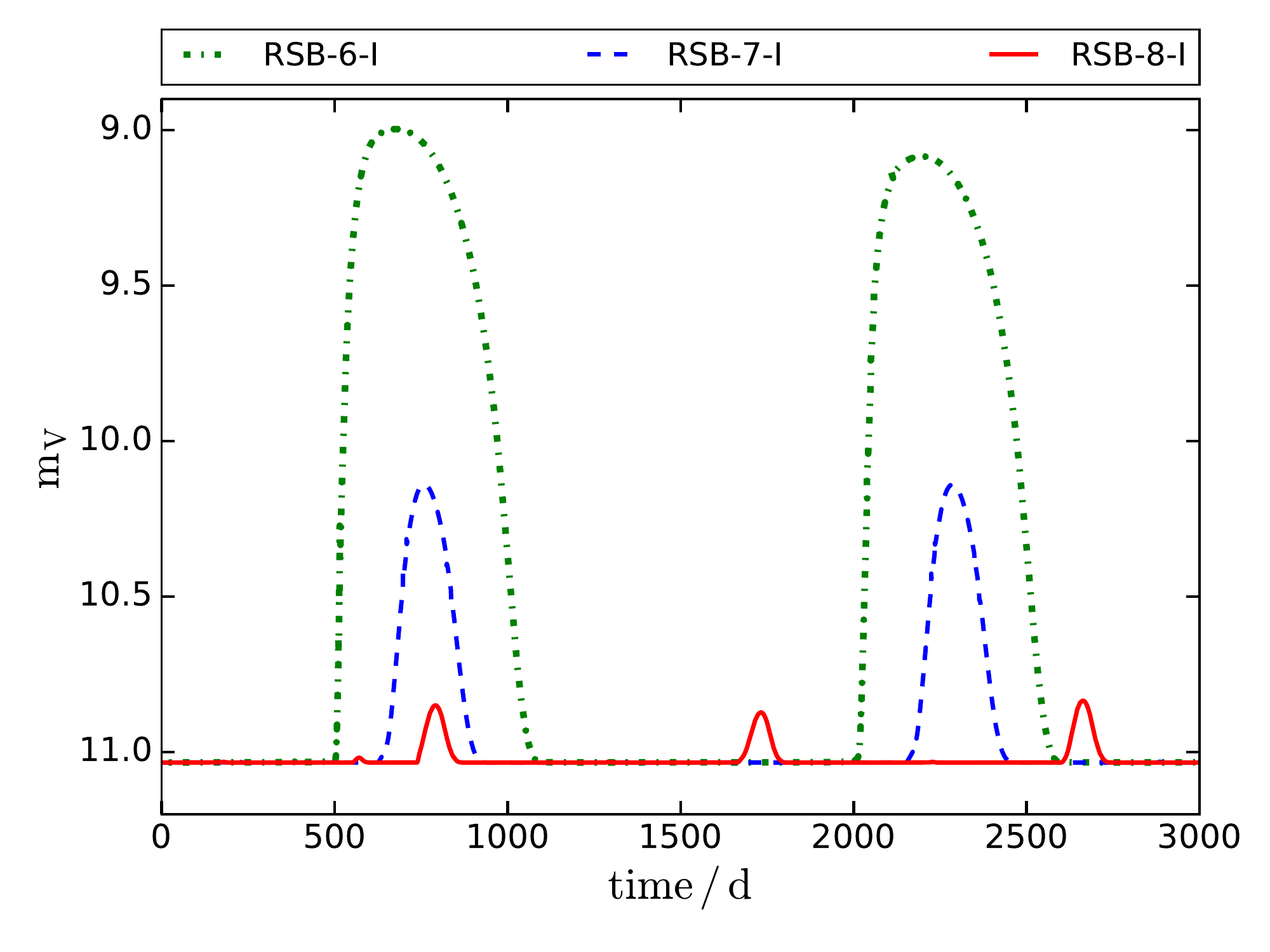}
 \caption{Optical light curves for irradiated models of RS~Oph -- RSB-6-I (dash-dotted green curve), RSB-7-I (dashed blue curve), RSB-8-I (solid red curve).}
\label{rsb_lc_irr}
\end{figure}

\begin{table*}
\caption{Disc outbursts properties for RS~Oph irradiated models.}
\begin{threeparttable}
\centering
\begin{tabular*}{2 \columnwidth }{@{\extracolsep{\fill}} l c c c c c c c c}
\hline
\hline \\ [-1.8ex]
      Model & 
      \multicolumn{1}{c}{$\dot{M}_{\rm in}$ (qsc)/}  &
      \multicolumn{1}{c}{$\dot{M}_{\rm in}$ (peak)/}  & 
      \multicolumn{1}{c}{$m_{\rm V}$}  &
      \multicolumn{1}{c}{$m_{\rm V}$}  &
      \multicolumn{1}{c}{$\Delta M$ /} &
      \multicolumn{1}{c}{$\tau_{\mathrm{rec}}$ /}  &
      \multicolumn{1}{c}{ $\tau_{\mathrm{ob}}$ /} &
      \multicolumn{1}{c}{ $R_{\mathrm{f,max}}$ /}\\
         &
        \multicolumn{1}{c}{ $\Msun$yr$^{-1}$} &
        \multicolumn{1}{c}{ $\Msun$yr$^{-1}$} &
        \multicolumn{1}{c}{(qsc)}  & 
        \multicolumn{1}{c}{(peak)} & 
        \multicolumn{1}{c}{ $\Msun$} & 
        \multicolumn{1}{c}{d} & 
        \multicolumn{1}{c}{ d}&
        \multicolumn{1}{c}{$10^{10}$ cm} \\ [0.8ex]
   
\hline \\ [-1.8ex]
RSB-6-I&$1.25 \times 10^{-12}$&$4.05 \times 10^{-6}$  & 11.03 &9.09&$5.08 \times 10^{-6}$ &1340& 538 &72.7\\
RSB-7-I&$1.45 \times 10^{-12}$&$1.17 \times 10^{-6}$ & 11.03 & 10.15 & 3.9$\times 10^{-7}$&1530& 305&43\\
RSB-8-I&$1.44\times 10^{-12}$&  $1.1-2.4\times 10^{-7}$ &11.03 & 10.85 & $1.5-3.9 \times 10^{-8}$& 935-1335 & 135-140 &16.8-22.5\\
\hline
\end{tabular*}
\begin{tablenotes}
\item Columns are the same as in Table \ref{tab:table_RSDIM}
\end{tablenotes}
\end{threeparttable}
\label{tab:table_RSIrr}
\end{table*}

\subsection{\textsl{Self-irradiated discs}}
We note from the results of the previous section that the accretion rate at inner edge gets quite high during outbursts, and can therefore lead to significant irradiation effects in the outer parts of the accretion disc (see Eq.~\ref{eq:flratio}), because the white dwarf is massive. Here we consider irradiation using Eq.~\ref{eq:flirrad} for the entire disc. This is appropriate because irradiation effects are important only in the outer disc. We consider similar models to the three that we discussed previously -- RSB-6-I, RSB-7-I, RSB-8-I --  with the irradiation coefficient ${\cal C} = 0.005$. Results are shown in Table. \ref{tab:table_RSIrr}, and the light curves (V magnitude) are shown in Fig. \ref{rsb_lc_irr}. 
\begin{description}

\item{-- RSB-6-I:} $\dot{M}_{\rm in}$ at maximum is almost the same as in the unirradiated case, but the outburst lasts longer and the mass deposited on the WD during an outburst is five times larger. The recurrence time is 2.5 times longer than for RSB-6, but this is still not consistent with the observations; roughly 5 outbursts are expected to occur in a 20 yr span, and each outburst can trigger a thermonuclear runaway. As before, one can therefore reject this model.
 
\item{-- RSB-7-I:} Irradiation has a significant impact in this model. The $\dot{M}_{\rm in}$ peak is roughly 1.85 times that obtained for RSB-7. The mass deposited on the WD during an outburst is very close to $M_{\rm ign}$ and could trigger a recurrent nova, in contrast to RSB-7 for which at least 6 outbursts were required to accrete the same mass. The recurrence time is 4.5 times longer than for RSB-7, but because of the increased accumulated mass during a single outburst, roughly 4 to 5 outbursts are expected to occur in a 20 yr span. Each outburst can again trigger a thermonuclear runaway. For this reason, this model also conflicts with the observations.

\item{-- RSB-8-I:} Irradiation effects are the strongest of the three models we considered. Here, we get a cycle of slightly different outbursts, hence the range given in table \ref{tab:table_RSIrr} for this model. These outburst parameters, however, do not vary by more than a factors 2. Both the peak accretion rate and mass accreted during an outburst are an order of magnitude larger than in the case of RSB-8, but the mass accreted during a single outburst is clearly less than the ignition mass. Now 18 outbursts are required to accrete the ignition mass, and a recurrent nova can be triggered once in every 55 yr. Although more mass is accreted when irradiation is taken into account, the recurrence time is also significantly increased and the time required to trigger a recurrent nova is similar to RSB-8 case. 

\end{description}
We therefore conclude that, as for the unirradiated case, that if dwarf-nova outburst are to be recurrent nova triggers, the mass-transfer rate must lie in the range $10^{-7}$ to $10^{-8} ~\Msun$~yr$^{-1}$. This is therefore a general conclusion, independent of the magnitude of the irradiation effect.

\subsection{Does one need dwarf novae to trigger RS~Oph TNR outbursts?}

Finally, we should check if a more conventional, not involving dwarf-nova outbursts,  scenario can explain the RS~Oph recurrent nova outbursts. A steady accretion rate of $2\times 10^{-8}\Msun$~yr$^{-1}$ is sufficient to accumulate  the mass $M_{\rm ign}$ corresponding to the RS~Oph parameters. A disc in a system with RS~Oph binary parameters is stable if the rate at which it is fed mass satisfies the inequality $\dot M_{\rm tr}> \dot{M}^{+}_{\rm crit}$, where the stable hot-disc minimal accretion rate is (Eq.~\ref{eq:Mdotplusapp})
\begin{equation}
\dot{M}^{+} = 1.44 \times 10^{-10} \alpha_{\rm h}^{-0.008} m_{\rm wd}^{-0.88}R_{\rm D,10}^{2.67}\,\Msuny;
\label{eq:Mdotplus}
\end{equation}
$m_{\rm wd}$ is the white-dwarf mass in solar units, and $R_{\rm D,10}$  the outer disc radius in units of $10^{10}$cm. For $\dot M_{\rm tr}> 2\times 10^{-8}\Msun$~yr$^{-1}$ an accretion disc with an outer radius $R_D < 7.4 \times 10^{10}$cm is stable with respect to the thermal-tidal instability. This is however quite small, but marginally possible if accretion occurs via a wind (the circularization radius must be  smaller than the disc size).
In such a scenario there would be no direct explanation for the low accretion rate deduced from \textsl{ROSAT} and \textsl{ASCA} observations and one would be left with speculations of type presented by \citet{Mukai2008}, who suggested that RS~Oph might be a fast rotator with essentially no boundary layer. Note that this is inconsistent with our assumption of a significant magnetic field.

\subsection{\textsl{GAIA} revised distance}

We finally comment on the recently updated distance of $2.26 \pm .27$ kpc by \textsl{GAIA} \citep{gaiaDR2}, a 40 percent increase with respect to the distance we have used here. With this new distance, the optical magnitude of the entire light curve would increase by 0.7 mag, all parameters being kept at their previous value. In order to still account for the observed quiescent luminosity, the secondary temperature has to be increased to 3800 K; the outburst amplitude is then reduced by a factor less than 2, all other parameters being unaltered. The (undetected) dwarf nova outbursts of RS~Oph would thus be slightly fainter, but the total mass accreted on to the white dwarf surface as well as the sequence of dwarf nova outbursts and nova eruptions remains exactly the same as previously and our conclusions for RS~Oph remain unaltered.
\begin{figure*}
\begin{center}
\includegraphics[width=\textwidth]{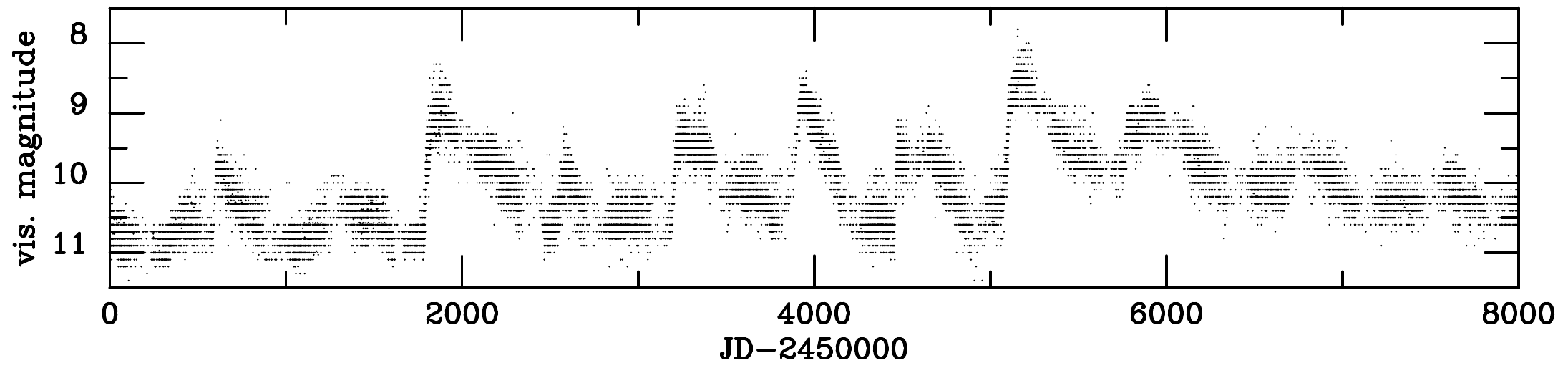}
\caption{Z~And visual light curve (data from AAVSO)}
\label{fig:lc_aavso_zand}
\end{center}
\end{figure*}

\section{Z~Andromedae}

Z~And is a classical symbiotic star consisting of an $M_1=0.65 \pm0.28~M_{\odot}$ \citep{SS1997} white dwarf, and a giant secondary with mass ratio $M_2/M_1 \approx 2.0$. Its orbital period is 759 d \citep{FL1994}, the inclination is  $59_{-2}^{+1 \; \circ}$ \citep{skopal2012}. Until recently, the distance to Z~And was estimated to be $1.2\pm0.5$ kpc \citep{Kenny1995}, while the recent parallax measured by \textsl{GAIA} \citep{gaiaDR2} puts Z~And at a distance of $1.95\pm0.078$ kpc. As for RS~Oph, we use here the old value of 1.2 kpc for comparison with previous studies, and comment at the end of this section on the impact of increasing the distance by 70 percent. The presence of a magnetized ($10^4$ to $10^{6}$ G) primary is deduced from observations of persistent optical pulsations at a period of 28 minutes \citep{sb1999}. 

As in most classical symbiotic stars, the white dwarf in Z~And (the hot component) exhibits a quasi-steady luminosity component of about $10^3~\Lsun$ attributed to thermonuclear shell burning of hydrogen at rate of few times $10^{-8}\Msun$~yr$^{-1}$ (see, e.g., S2006). Symbiotic stars undergo outbursts, the most common of which are \textsl{classical symbiotic outbursts} the origin of which is unknown. Their high peak luminosities exclude dwarf-nova type outbursts while their high repetition frequency rules out thermonuclear runaways of the recurrent-nova type. Contrary to the recurrent novae, white dwarfs in these symbiotic systems have low masses, precluding short recurrence times.

Fig. \ref{fig:lc_aavso_zand} shows the visual light curve of Z~And from 1995 to 2017, obtained using AAVSO data. The light curve is complex, and displays a variety of outbursts of different amplitudes and durations, which are not necessarily caused by the same mechanism.

In 1997, the Z~And light-curve showed a low amplitude (about 1 mag) outburst in the \textsl{V}-light, with a rise-time shorter than about 20~d followed by a decay lasting about 200~d (first outburst in Fig. \ref{fig:lc_aavso_zand}). The asymmetric outburst shape prompted S2006 to suggest that the 1997 outburst was due to a dwarf-nova type instability of the outside-in type, i.e. with an outburst starting in the outer disc regions \citep[see, e.g.,][]{BM01}. A more powerful outburst was observed in 2000 to 2002, when after a rise similar to the 1997 event the brightness kept increasing to reach a 2 optical mag amplitude, corresponding to a bolometric luminosity of about $10^4\,\Lsun$ (second outburst in Fig. \ref{fig:lc_aavso_zand}, starting at JD$\approx$ 2451800). If accretion-powered, this high-state would correspond to an accretion rate $\dot M \gtrsim 10^{-5}\Msun$~yr$^{-1}$. Such a high accretion rate would have to be sustained for a full year during which the system maintained a luminosity close to its maximum (S2006). Since this seems unlikely, S2006 proposed that this event is a `combination nova' outburst in which a dwarf-nova outburst triggered increased shell nuclear burning. In order to test S2006 hypotheses we calculate models of disc outbursts in Z~And using our DIM code, adapted to the physical properties of  this system.

\begin{table*}
\caption{Outburst properties of Z~And non-irradiated disc models.}
\begin{threeparttable}
\centering
\begin{tabular*}{\textwidth }{@{\extracolsep{\fill}} l c c c c c c c c c}
\hline
\hline \\ [-1.8ex]
      Model & 
       \multicolumn{1}{c}{$\dot{M}_{\mathrm{in}}$ (qsc)/}  &
      \multicolumn{1}{c}{$\dot{M}_{\rm in}$(peak)/}& 
      \multicolumn{1}{c}{$\Delta m_{\rm V}$}  &
      \multicolumn{1}{c}{$\Delta M$ /}  &
      \multicolumn{1}{c}{$R_{\mathrm{trig}}$ /}&
       \multicolumn{1}{c}{$R_{\mathrm{f,max}}$ /}&
      \multicolumn{1}{c}{$\tau_{\mathrm{rise}}$ /}  &
      \multicolumn{1}{c}{$\tau_{\mathrm{dec}}$ /}  &
      \multicolumn{1}{c}{$\tau_{\mathrm{rec}}$ /}  \\
         &
         \multicolumn{1}{c}{ $\Msuny$} &
        \multicolumn{1}{c}{$\Msuny$} &
       \multicolumn{1}{c}{                    }  &  
        \multicolumn{1}{c}{$\Msun$} & 
        \multicolumn{1}{c}{$10^{10}$ cm} &  
      \multicolumn{1}{c}{ $10^{10}$ cm} &   
       \multicolumn{1}{c}{\rm d} & 
      \multicolumn{1}{c}{ \rm d} & 
        \multicolumn{1}{c}{\rm d}\\ [0.8ex]
        
\hline \\ [-1.8ex]
ZAB-6&$3.3 \times 10^{-11}$&$3.4 \times 10^{-6}$  & 0.8 & 5.6 $\times 10^{-7}$ & 4.05 & 40  & 73 & 118 & 355\\
ZAB-7&$3.3 \times 10^{-11}$&$5.2 \times 10^{-7}$ & 0.2 & 4.0 $\times 10^{-8}$ & 2.15 & 19 &30 & 46 & 208 \\
ZAB-8&$3.3 \times 10^{-11}$ &$8.4 \times 10^{-8}$ & 0.05 &$3.1\times 10^{-9}$  & 1.5  & 10 & 14 & 18 & 143 \\
\hline
\end{tabular*}
\begin{tablenotes}
\item The first column lists the models, the second column gives the value of the accretion rate during quiescence, the third column gives the accretion rate attained during the peak of the outburst. The fourth column gives the outburst amplitude in visual magnitudes. The next column contains an estimate of the amount of mass accreted on to the white dwarf during the outburst. The following two columns provide respectively the radii at which the heating front starts and finishes its propagation. The next two columns give the outburst rise and decay times observed in V. The last column gives the recurrence time between outbursts. \end{tablenotes}
\end{threeparttable}
\label{tab:tablenoirr}
\end{table*}

\subsection{Dwarf nova outbursts of Z~And}

As in the case of RS~Oph, it is not clear how in Z~And the mass lost by the giant companion forms an accretion disc around the white dwarf. 
The orbital separation of \Zand is $a=3.053 \times 10^{13}$ cm, to which corresponds a circularization radius $R_{\rm circ}=2.085 \times 10^{12}$ cm.
In the case of wind accretion the circularization radius would be $1.34\times 10^{10}$cm, assuming that the wind speed is equal to the escape speed from the giant, $\lambda(a) \approx 1$.

Since the critical accretion rate at $R_{\rm circ}$ is (Eq.~\ref{eq:Mdotplusapp}) $2.95 \times 10^{-4}\,\Msuny$, a non-irradiated disc in Z~And is thermally and viscously unstable for all realistic (less than $10^{-4}\,\Msuny$) mass-transfer rates. However, outbursts of \Zand should differ in several aspects from those observed in normal dwarf-novae. 

First, the presumed radius of the accretion disc in Z~And, $R_D\approx 1.4 - 2.8 \times 10^{12}$ cm (S2006), is one or two orders of magnitude larger than that in dwarf-novae which, as already noticed by \citet{duschl1986b}, can have dramatic effects on heating and cooling front propagation \citep[see e.g.,][and  Section \ref{subsect: dnrsoph} about RS~Oph]{Menou2000,dubus2001}. 

Secondly, the disc is irradiated by a constant white dwarf surface luminosity of about $10^3\,\Lsun$, corresponding to an effective temperature of $150\,000$\,K, while temperatures of white dwarfs in dwarf-novae are an order of magnitude lower \citep[see e.g.,][]{Urban2000}. Finally the observed visual light-curves have a major contribution from the giant companion \citep{duschl1986b,kenyon1986} and the duration of the outbursts is shorter than the orbital period which might provide an additional source of variability.

Thirdly, according to S2006, the asymmetric shape of the 1997 outburst light-curve suggests an \textsl{outside-in} event  (i.e. an outburst in which the heating front propagates inwards from the outer disc regions), if interpreted as a dwarf-nova outburst \citep{S84a}. One should note, however, that `fast-rise -- slow-decay' light-curves can be also produced by outbursts starting near the inner disc radius. This happens e.g., in the case of truncated and/or irradiated discs.  In addition {outside-in} outbursts  are unlikely to occur in very large discs\footnote{Model outbursts for LMXB parameters are always of the inside-out type \citep{dubus2001}}. The criterion for an outside-in outburst can be written as \citep{Lasota2001}
\begin{equation}
\label{eq:o-i}
\dot M_{\rm tr} \gtrsim \frac{0.5}{\sqrt{\delta}}\dot M^-_{\rm crit},
\end{equation}
where $1 < \delta \lesssim 2$,
\begin{equation} 
\dot M^-_{\rm crit}= 1.15 \times 10^{-5} \alpha_{\rm c}^{-0.02} \left( \frac{M_1}{\Msun} \right)^{-0.88}R_{D,12}^{2.65}\,\Msuny
\end{equation}
is the critical (maximal)  accretion rate for a cold and stable accretion disc (Eq.~\ref{eq:Mdotminus}), and $R_{\rm D,12}$ is the disc radius in 10$^{12}$~cm units. For the Z~And parameters Eq.~(\ref{eq:o-i}) implies rather unrealistically  high mass-transfer rates, larger than about $10^{-5}\,\Msuny$.

\subsubsection{Non-irradiated disc}

We will first check the S2006 suggestion that the disc instability model can explain the asymmetrical, 1 mag feature observed in 1997 in the visual light-curve of Z~And.
In order to separate and evaluate the various components contributing to the light-curve of a hypothetical dwarf-nova type outburst of Z~And, we first consider the case of a non-irradiated disc. 
We assume a white dwarf magnetic moment of $5.12\times 10^{31}$ G${\rm ~cm}^3$, which corresponds to a magnetic field close to $10^5$G. Our results do not depend significantly on the accretor's magnetic moment. 

In calculating the visual light-curves we include the contribution of the giant companion in a different way from \cite{shl2003}. Their approach is well-adapted to a Roche-lobe filling low-mass companion but not applicable to the case of \Zand.
We considered each of the two extreme simplified assumptions.\\
 a)  The secondary luminosity does not change in response to irradiation. This is appropriate if for example the irradiated side is not facing the observer. The secondary contribution is fixed to be 10.8 mag.\\
 b) The response of the secondary luminosity to irradiation is maximal. This would be appropriate if the irradiated side of the secondary is facing the observer, and if the secondary radius is as large as possible.
In this case we use the minimum acceptable secondary effective temperature, which we take to be 2900~K and estimate the radius that would correspond to the {\Zand} quiescent visual magnitude (10.5 mag); the secondary radius is then found to be 110 $\Rsun$, so the secondary fills 57 per cent of its Roche lobe.

We found that the two prescriptions produce very similar light-curves and in what follows we present light-curves calculated using assumption (b) only.

We consider outburst models for three mass-transfer rates: ZAB-8, ZAB-7 and ZAB-6 for respectively $\dot M_{\rm tr}= 10^{-8}\,,10^{-7}$ and $10^{-6}\,\Msuny$.
We assume $\alpha_c=0.02$ and $\alpha_h=0.1$, and chose $c_{\rm tid}$ so that the outer disk radius is $R_{\rm D} = 2.37\times 10^{12}$cm. Following S2006 we assume a distance of 1.2 kpc, for the purpose of comparing our results with their work, although we shall later comment on the recently updated distances by \textsl{GAIA}.

\begin{figure}
\includegraphics[width=\columnwidth]{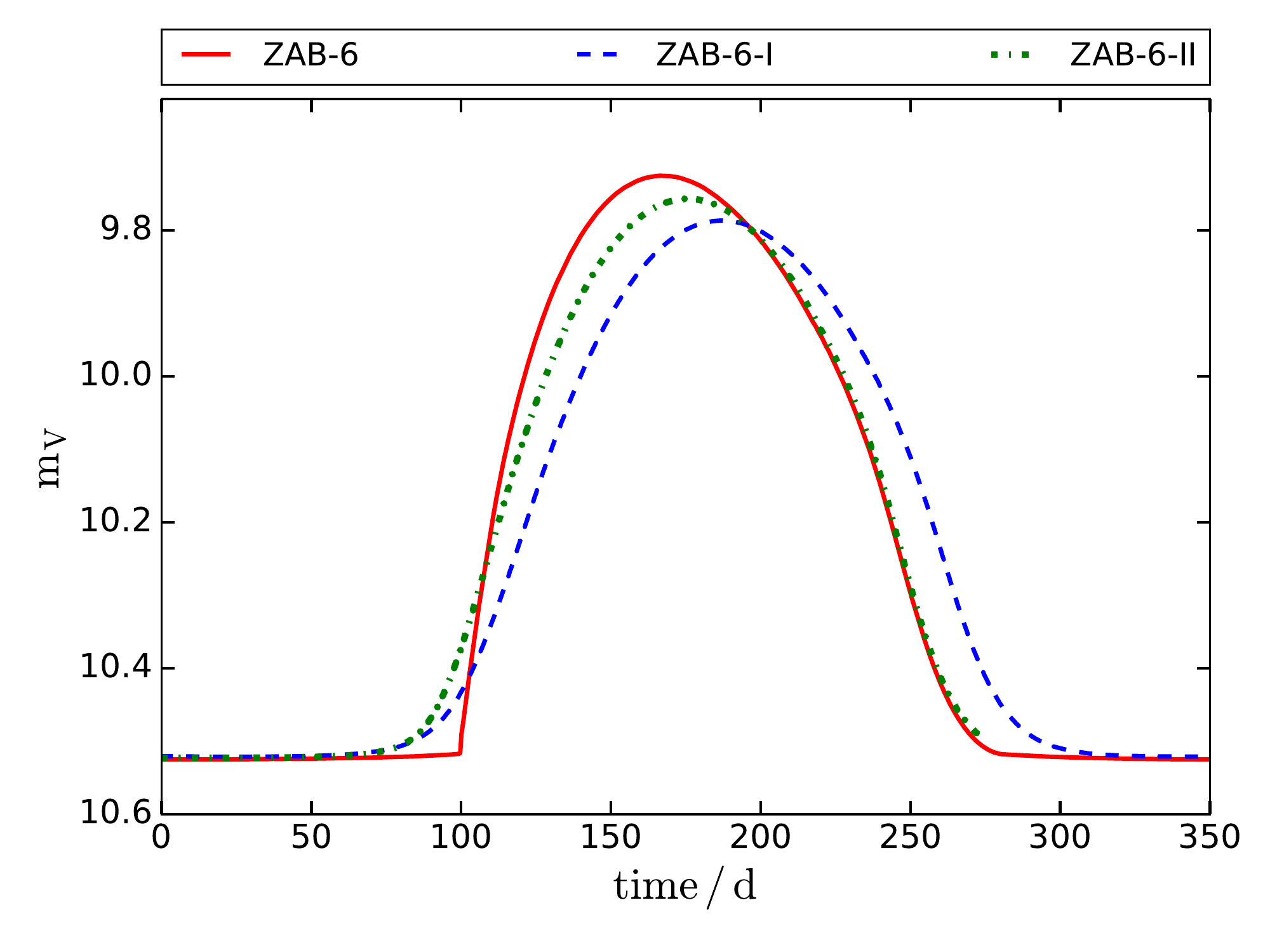}
\caption{Visual light-curve of dwarf-nova outbursts for the models ZAB-6 (solid red curve), ZAB-6-I (dashed blue curve) and ZAB-6-II (dash-dotted green curve). The properties of the light-curves are given in Tables \ref{tab:tablenoirr} and \ref{tab:tableirr}.}
\label{fig:noirr-lc}
\end{figure}

The properties of these three models are presented in \autoref{tab:tablenoirr}. One can see that only model ZAB-6 produces a visual amplitude not too different from 1 mag; however, the shape of the outburst light-curve
is markedly different from that observed in 1997 in \Zand. The rise-time (73~d) is three times longer, while the decay-time (118~d) is roughly half that observed. The light-curve shown in Fig. \ref{fig:noirr-lc}
is typical of an inside-out outburst, i.e. of an outburst starting in the inner disc region. Indeed, the inner disc radius during quiescence, given by Eq.~(\ref{eq:rmag}) with $\mu_{30}=51.2$ and $\dot M_{\rm acc}=3.3 \times 10^{-11}\Msuny$ is equal to $R_{\rm in}=0.82\times 10^{10}$cm whereas the heating front starts propagating at $R_{\rm trig}=4.05\times 10^{10}$cm. While the inner disc edge reaches the white dwarf surface, this front propagates out up to $R_{\mathrm{f,max}}=40 \times 10^{10}$cm, far from the outer disc rim. The outermost disc regions are not affected by the fronts and the outer disc radius remains fixed because the amount of angular momentum deposited by the outburst is too small to affect it in a significant way.
\begin{figure}
\includegraphics[width=\columnwidth]{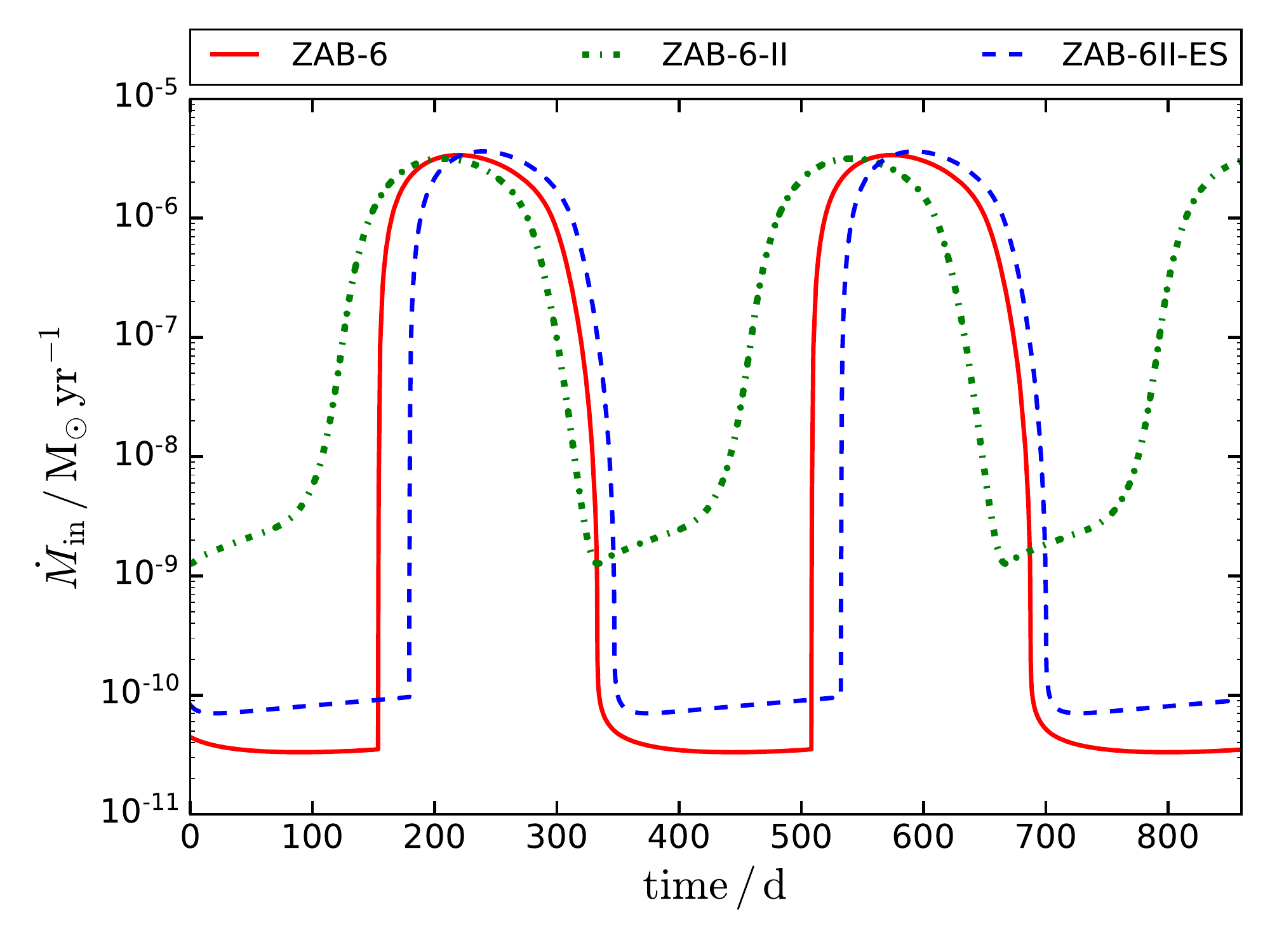}
\caption{Accretion rate at the inner disc edge during outburst for models ZAB-6 (solid red curve), ZAB-6-II (dash-dotted green curve) and ZAB-6II-ES (dashed blue curve).}
\label{fig:mdotin}
\end{figure}
Although the visual amplitude is small, the accretion rate at the inner edge increases by several orders of magnitude, by a factor of about $10^5$ in the case of model ZAB-6 (see Fig. \ref{fig:mdotin}). At maximum the inner edge is squeezed by accretion on to the white dwarf surface (the white dwarf radius $R_{\rm WD}=8.56 \times 10^8$cm) and for this model the corresponding accretion luminosity reaches $2600\,\Lsun$, which should be emitted mostly in EUV. This is consistent with the observed increase of the white dwarf temperature from $150\,000$\,K to $180\,000$\,K during the 1997 outburst. The mass dropped on to the white dwarf during a ZAB-6 outburst is $5.6 \times 10^{-7}\,\Msun$. 

Since models ZAB-7 and ZAB-8 reproduce neither the shape nor the amplitude observed during the 1997 outburst, the only potential candidate for a model explanation of this event is ZAB-6, corresponding to an accretion rate of $10^{-6}\,\Msun$, two to three orders of magnitude larger than the mass-transfer rate considered to be typical of systems such as {\Zand} (S2006). Therefore the 1997 outburst would have to be related to an increase in the mass-transfer rate and could not be a pure dwarf-nova outburst. It would be similar to dwarf-nova \textsl{superoutbursts} according to the interpretation of \citet{HLW00} and  \citet{Smak08}.

But the {\Zand} accretion disc is irradiated and its structure can be substantially affected by irradiation. We discuss the impact of irradiation in the next subsection.
\begin{table*}
\caption{Outburst properties of Z~And irradiated disc models. I-- fully irradiated discs.}
\begin{threeparttable}
\begin{tabular*}{\textwidth }{@{\extracolsep{\fill}} l c c c c c c c c c c }
\hline
\hline \\ [-1.8ex]
      Model & 
      \multicolumn{1}{c}{$\dot{M}_{\rm in} (\rm qsc)$} & 
      \multicolumn{1}{c}{$\dot{M}_{\rm in} (\rm peak)$} & 
      \multicolumn{1}{c}{$\Delta m_{\rm V}$} & 
      \multicolumn{1}{c}{$\Delta M$} & 
       \multicolumn{1}{c}{$R_{\mathrm{in, qsc}}$}&
      \multicolumn{1}{c}{$R_{\mathrm{f,min}}$}&
       \multicolumn{1}{c}{$R_{\mathrm{f,max}}$}&
      \multicolumn{1}{c}{$\tau_{\rm rise}$}  &
      \multicolumn{1}{c}{$\tau_{dec}$} &
      \multicolumn{1}{c}{$\tau_{\rm rec}$} \\
         &
        \multicolumn{1}{c}{/ $\Msuny$} &
        \multicolumn{1}{c}{/ $\Msuny$} &        
   \multicolumn{1}{c}{} & 
       \multicolumn{1}{c}{/ $\Msun$)} &
        \multicolumn{1}{c}{/ $10^{10}$ cm} &  
      \multicolumn{1}{c}{/ $10^{10}$ cm} &  
       \multicolumn{1}{c}{/ $10^{10}$ cm} & 
       \multicolumn{1}{c}{/ d} &
       \multicolumn{1}{c}{/ d} &
       \multicolumn{1}{c}{/ d} \\ [0.8ex]
\hline \\ [-1.8ex]
ZAB-6-I & $2.96 \times 10^{-9}$&$2.98\times 10^{-6}$& 0.7 & $4.63\times 10^{-7}$& 0.24 & 3.3 &40.2 &122 &123 & 335 \\
ZAB-7-I &$3.04 \times 10^{-9}$& $2.65\times 10^{-7}$ & 0.13 &  $1.51\times 10^{-8}$ & 0.23 &3.6&16.1 &65 &57 & 129 \\
ZAB-8-I &--&-- &-- &-- &-- &--&--&--&-- & --\\
ZAB-6-II &$2.02\times 10^{-9}$&$3.16\times 10^{-6}$&0.76 &  $4.98\times 10^{-7}$ & 0.29 &2.5 &40.5&110 &116 & 335 \\
ZAB-7-II &$1.61\times 10^{-9}$&$3.44\times 10^{-7}$&0.16& $2.11\times 10^{-8}$  & 0.29 &2.5 &17.0&55 &52 & 142 \\
ZAB-8-II &$2.05\times 10^{-9}$&$2.17\times 10^{-8}$& 0.0017& $4.02\times 10^{-10}$ & 0.25 & 2.9 & 6.7&20 &20 & 44 \\
\hline
\end{tabular*}
\begin{tablenotes}
\item Columns are the same as in Table \ref{tab:tablenoirr}
\end{tablenotes}
\end{threeparttable}
\label{tab:tableirr}

\end{table*}

\subsubsection{Irradiated discs}

In this section we consider irradiation, under two different assumptions for the irradiation efficiency at the outer disc.

\paragraph{Fully irradiated discs.}
\begin{figure}
\includegraphics[width=\columnwidth]{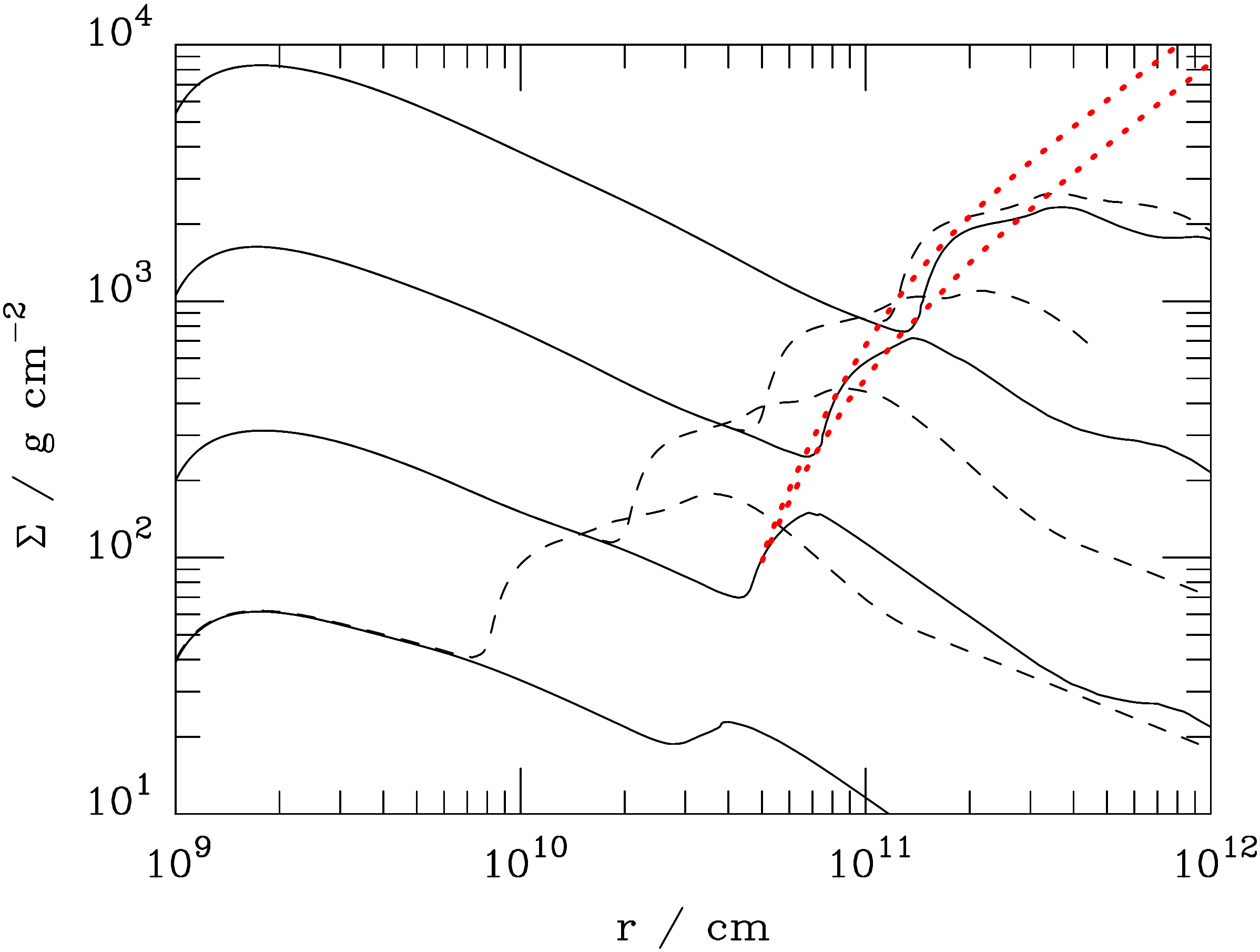}
\caption{Column density ($\Sigma(R)$) profiles of stationary discs around a $0.65\Msun$ white dwarf irradiated by a constant luminosity $L= 4\times 10^{36}\ergs$.  Disc irradiation is described by Eq.~\ref{eq:flirrad}, with parameters ${\cal C} = 0.002$ and $\alpha=0.1$. The accretion rates from bottom to top are $10^{16}$, $10^{17}$, $10^{18}$, and $10^{19}$ g~s$^{-1}$. Dashed lines represent non-irradiated discs, full lines correspond to disc irradiated by a constant luminosity.
The position of the critical column densities $\Sigma^+$ and $\Sigma^-$ are marked by two dotted red lines intersecting on the diagram at $r = 5\times 10^{10}$cm,  $\Sigma = 100\,\rm g\;cm^{-2}$.}
\label{fig:radsigma}
\end{figure}

We first consider the case where irradiation by both nuclear burning at the white dwarf surface and accretion modify the structure of the entire disc (i.e. both inner and outer regions). 
As mentioned in \autoref{sec:irrad}, when the disc scale-height $H$ is such that $H/R_{\rm WD}< 1$, we  use the extended source formula given by Eq.~(\ref{eq:fliradwd}), with $T_{\rm irr, S}^4 = T_*^4+T_{\rm eff}^4$ \citep{HLD99}.
For $H/R_{\rm WD}> 1$ we use  Eq.~(\ref{eq:flirrad}) with $L=L_* + L_{\rm acc}$.

We consider the same three mass-transfer rates as before, but with two sets of irradiation parameters, a strong irradiation case, ZAB-6-I, ZAB-7-I and ZAB-8-I with $1-\beta=0.03$ and ${\cal C}=0.002$, and a weaker irradiation case, ZAB-6-II, ZAB-7-II and ZAB-8-II with 
$1-\beta=0.01$ and ${\cal C}=0.001$. Since we do not have direct (or even indirect) information about disc irradiation in Z~And or similar systems we are using parameters similar to those used in modelling dwarf-novae and X-ray transient systems. There is no guarantee that this choice is adequate but, as we shall see, the results are rather insensitive to irradiation parameters, at least as far as the main properties of the light-curves are concerned.

Constant irradiation of the outer disc has a different effect on disc stability than accretion self-irradiation the luminosity of which is proportional to the accretion rate. Comparing Fig. \ref{fig:radsigma} with figs. 8 and 9 of \citet[]{dubus1999}, one can see that in both cases irradiation extends the radius of the hot stable portion of the disc, but in the case of constant irradiation this effect increases with decreasing accretion rate so that discs with accretion rates less than about $10^{17}$
g~s$^{-1}$ are fully \textsl{thermally} stable. Our model ZAB-8-I  (corresponding to accretion rate $6.3\times 10^{17}$g~s$^{-1}$) is locally unstable but because of the stabilizing effects of the radial terms in the time-dependent energy-conservation equation \citep[the term proportional to $J$ in equation 7 of][]{dubus2001} no outburst is produced. 

\begin{figure}
\includegraphics[width=1.\columnwidth]{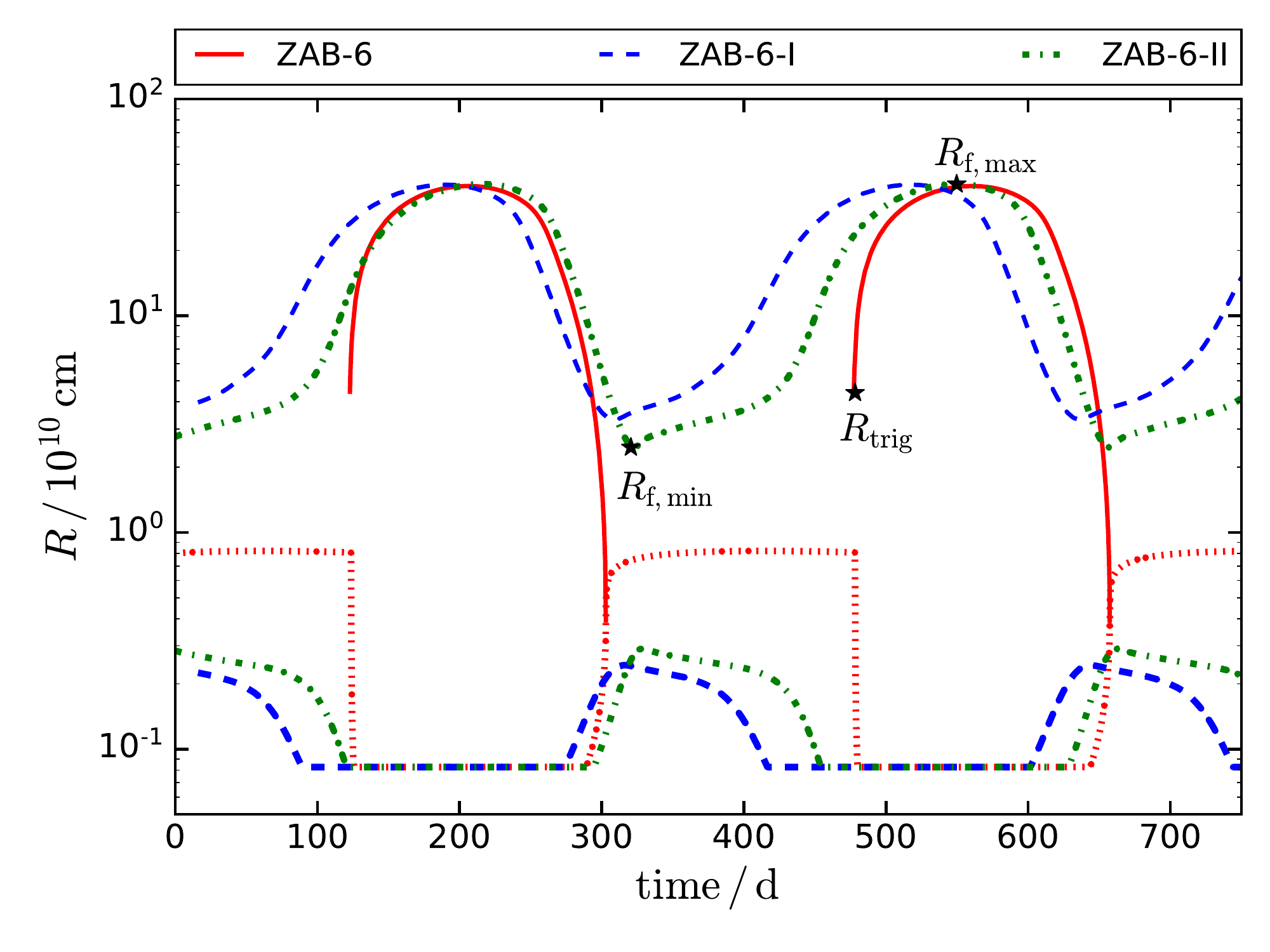}
\caption{Position of heating / cooling fronts during outbursts for models ZAB-6, ZAB-6-I and ZAB-6-II (top red solid, blue dashed and green dash-dotted curves respectively). Positions of the inner disk radius during the outburst for these models are also shown (bottom curves), using the same line-styles except for ZAB-6 model (red dotted curve). The radius at which the instability is triggered, $R_{\rm trig}$, is marked for the second of the ZAB-6 outbursts. Similarly, the maximum radius $R_{\rm f,max}$ reached by the heating front and the minimum radius $R_{\rm f,min}$ reached by the cooling front are indicated for the second of the ZAB-6-II outbursts.}
\label{fig:rfronts}
\end{figure}

\begin{table*}
\caption{Outburst properties of Z~And irradiated disc models. II -- inner irradiation only. }
\begin{threeparttable}
\begin{tabular*}{\textwidth }{@{\extracolsep{\fill}} l c c c c c c c c c c}
\hline
\hline \\ [-1.8ex]
      Model & 
      \multicolumn{1}{c}{$\dot{M}_{\rm in} (\rm qsc)$} & 
      \multicolumn{1}{c}{$\dot{M}_{\rm in} (\rm peak)$} & 
      \multicolumn{1}{c}{$\Delta m_{\rm V}$} & 
      \multicolumn{1}{c}{$\Delta M$} & 
       \multicolumn{1}{c}{$R_{\mathrm{in, qsc}}$}&
      \multicolumn{1}{c}{$R_{\mathrm{trig}}$}&
       \multicolumn{1}{c}{$R_{\mathrm{f,max}}$}&
      \multicolumn{1}{c}{$\tau_{\rm rise}$}  &
      \multicolumn{1}{c}{$\tau_{dec}$} &
      \multicolumn{1}{c}{$\tau_{\rm rec}$} \\
         &
        \multicolumn{1}{c}{/ $\Msuny$} &
        \multicolumn{1}{c}{/ $\Msuny$} &        
   \multicolumn{1}{c}{} & 
       \multicolumn{1}{c}{/ $\Msun$)} &
        \multicolumn{1}{c}{/ $10^{10}$ cm} &  
      \multicolumn{1}{c}{/ $10^{10}$ cm} &  
       \multicolumn{1}{c}{/ $10^{10}$ cm} & 
       \multicolumn{1}{c}{/ d} &
       \multicolumn{1}{c}{/ d} &
       \multicolumn{1}{c}{/ d} \\ [0.8ex]
\hline \\ [-1.8ex]
ZAB-6-II-ES & $7.0 \times 10^{-11}$&$3.63\times 10^{-6}$& 0.84 & $5.9\times 10^{-7}$& 0.67 & 4.17 &40.2 &64 &108 & 352\\
ZAB-7-II-ES &$7.0 \times 10^{-11}$& $5.45\times 10^{-7}$ & 0.22 &  $4.17\times 10^{-8}$ & 0.66 &2.28&18.8 &28 &44 & 212 \\
ZAB-8-II-ES& $7.0 \times 10^{-11}$ &$8.11\times 10^{-8}$ & 0.52 & $2.76\times 10^{-9}$  & 0.66 & 0.87 &8.8 &13&17 & 139 \\
\hline
\end{tabular*}
\begin{tablenotes}
\item Columns are the same as in Table \ref{tab:tablenoirr}
\end{tablenotes}
\end{threeparttable}
\label{tab:tableirrin}
\end{table*}

Outbursts in a disc irradiated by a constant luminosity source (which dominates except at maximum in the ZAB-6 models) are different from those occurring in non-irradiated discs or in self-irradiated discs of soft X-ray transient sources. They consist of reflections of heating and cooling fronts travelling between two radii defined as $R_{\mathrm{f,min}}$ and $R_{\mathrm{f,max}}$. This can be seen in Fig.~\ref{fig:rfronts}, for ZAB-6-I and ZAB-6-II, where the fronts propagate between  $R_{\mathrm{f,max}}$ and $R_{\mathrm{f,min}}$, so that the cooling fronts never reach the inner disc radius shown by dashed lines. This is a major difference with ZAB-6 in which the cooling front can propagate to the disc inner edge, and bring the full disc into quiescence. 
Due to irradiation the inner disc is kept hot during the whole outburst cycle which produces cooling front reflections whereas, as in the non-irradiated case, the heating front is stopped well before reaching the disc outer rim. Contrary to the case of soft X-ray transients \citep{dubus2001} disc truncation does not suppress reflares because irradiation is dominated by its constant component, independent of the accretion rate. 
Although the disc never fully enters into quiescence, the light curve shows clear, well separated outbursts, with quiescent intervals during which the optical light is dominated by the secondary and one is thus able to define the same parameters $\tau_{\rm rec}$, $\tau_{\rm rise}$, $\tau_{\rm dec}$ describing the outbursts properties as previously.

The resulting visual light-curves (see Fig. \ref{fig:noirr-lc} and \autoref{tab:tableirr} for the outburst parameters) for the highest mass-transfer rate case, $\dot M_{\rm tr}= 10^{-6}\,\Msuny$ are similar to those obtained in the non-irradiated  case but more symmetric. Since the heating fronts reach similar radii, the peak accretion rate is roughly the same as for non-irradiated discs and a comparable amount of mass is dropped by an outburst on to the white dwarf. The shape of the visual light-curve is even less similar to that observed during the 1997 outburst of Z~And. For lower mass-transfer rates, the outbursts we obtain are weaker than in the non-irradiated case and cannot really be qualified as outbursts. As before, to obtain the correct amplitude, the mass-transfer rate has to be increased by two orders of magnitude.

\paragraph{Inner disc irradiation only.} One could reasonably argue that describing outer disc irradiation in \Zand using the LMXB formalism adapted to the symbiotic case does not make much sense.
First, even in the case of LMXBs it is not clear how and why their outer discs are irradiated \citep{TMW90,dubus1999,tetarenko2018b}, even if observations clearly show that they are \citep[see e.g.,][]{VPMC95}. Secondly, in LMXBs the illuminating radiation is much harder than it is the case in \Zand. One could therefore argue that in the case of symbiotic systems the effect of outer disc irradiation is negligible.
This is why we have made another test of the hypothesis that some of symbiotic outbursts, those of \Zand in particular, are
dwarf nova outbursts or are triggered by them, using a different assumption about the irradiation effects. We now assume that irradiation is given by Eq.~(\ref{eq:fliradwd}) throughout the disc, whatever $H/R_*$, so that irradiation becomes very small for large radii. The irradiation flux now varies as $R^{-3}$ for large values of $R$.

As before we tried two cases:  `I',  with $1-\beta=0.03$, and `II' with  $1-\beta=0.01$ but since the results are almost indistinguishable we present in \autoref{tab:tableirrin} results only for ZAB-6-II-ES, ZAB-7-II-ES and ZAB-8-II-ES, where `ES' means that only inner disc irradiation is taken into account.

In contrast with the fully irradiated case, the cooling fronts are able to propagate to the inner disc radius; this is due to the fact that, during decay, the inner disc radius becomes large and the irradiation flux, when calculated using Eq.~\ref{eq:fliradwd}, becomes small. This  does not, however, lead to outbursts which differ in a very significant way from those obtained previously.

For the highest accretion rate ($10^{-6}\Msuny$) the model light-curves are very similar to that calculated for non-irradiated discs as can be seen in Figs. \ref{fig:noirr-lc} and \ref{fig:mdotin}. They are also
not much different from the light-curves obtained for a fully irradiated disc. This is not surprising because the importance of constant irradiation diminishes with increasing accretion rate and, because of the low white dwarf accretion efficiency, self-irradiation is (almost) never dominant. 

Differences in light-curves are more prominent for lower accretion rates, but these produce neither amplitudes nor shapes comparable to these observed.

\subsection{Combination-nova outbursts of Z~And}

As mentioned before, S2006 suggest that the 2000 to 2002 eruption of \Zand was powered by an increase in nuclear shell burning on the white dwarf surface triggered by mass influx added by a dwarf-nova outburst. They notice that this event seems to begin like
the 1997 outburst (that they assume to be a dwarf-nova outburst), but reaches a luminosity close to $10^4$ L$_\odot$ and lasts more than a year. In addition, observation of the ejection of an optically thick shell of material excludes a dwarf-nova outburst and points to a milder version of a nova event. S2006 propose that this was a disc-instability triggered thermonuclear eruption and call it `combination nova'. They estimate from observations that the mass dropped on to the white dwarf during the 1997 outburst was approximately $10^{-7}\,\Msun$. The mass of the burning shell was about $2\times 10^{-5}\,\Msun$ and S2006 consider that adding one to ten per cent of the shell mass should be sufficient to trigger a shell thermonuclear flash.

We do not address here the problems that such an interpretation poses for models of nuclear burning on the surface of a white dwarf mentioned by S2006 but consider only the viability of the dwarf-nova component of the combination nova. As we have shown above, a dwarf-nova outburst of a disc with \Zand parameters can provide more than about $10^{-7}\,\Msun$ only if the mass-transfer rate is close to $10^{-6}\,\Msuny$. Therefore such an outburst must be preceded by an increase of the mass-transfer rate by two orders of magnitude, since the long-term rate is supposed to be $10^{-8}\,\Msuny$. However, although such outbursts would provide the mass-increase necessary to trigger an enhanced thermonuclear burning shell on the white dwarf surface, the shape of their visual light-curves does not correspond well to observations. Therefore the hypothesis that the combination of dwarf-nova and nova outbursts is an explanation of \Zand symbiotic outbursts is not confirmed by model calculations of disc thermal-viscous instabilities. The fact that to get the right energetics requires an increase in the mass supply to the disc suggests that such an increase could be sufficient to trigger the outbursts of \Zand, as proposed by \citet{Leibowitz2008}. 

Using the updated distance from \textsl{GAIA} further supports this argument. With this new distance, the secondary has to be hotter than 3200 K for the quiescence magnitude to be around V=10.5, assuming that the secondary still fills 57 percent of its Roche lobe. The outburst amplitude is then reduced from 0.8 to 0.4 mag. If, on the other hand, we assume the temperature is 2900 K as before, then the secondary must entirely fills its Roche lobe, and the outburst amplitude is about 0.6 mag. In both cases, the dwarf-novae outbursts cannot explain the observed Z~And outbursts, pointing towards other possibilities such as mass-transfer enhancement as mentioned above.

\section{Conclusions}

We have studied the stability of accretion discs in symbiotic stars which usually have long orbital periods, with particular emphasis on two systems, RS~Oph and Z~And. For the expected mass-transfer rates lower than $10^{-4}\Msun$yr$^{-1}$, their presumably large accretion discs (larger than $10^{12}$ cm)  are prone to thermal-viscous instabilities and are thus expected to exhibit outbursts of the type observed in dwarf nova stars.

\subsection*{RS~Oph}
We have considered three models corresponding to three different mass-transfer rates, $10^{-6},~10^{-7}$ and $10^{-8}~\Msun$~yr$^{-1}$. None of these models reproduce the observed recurrence time and brightness amplitude of the recurrent nova outbursts. This rules out the possibility that these outbursts are in fact dwarf-nova type outbursts, contrary to the suggestion by \citet{KP09}, even in the case where their luminosity is increased due to steady hydrogen burning \citep{Alex2011}. Note also that the peak accretion rate found by \citet{Alex2011} exceeds the limit $\dot{M}_{\rm RG} = 3 \dot{M}_{\rm stable}$ above which hydrogen can be burned as quickly as it is accreted; their outburst luminosity is therefore largely overestimated and thermonuclear runaways are unavoidable.

We also considered the possibility that the disc instability outbursts can trigger a thermonuclear runaway on the white dwarf surface; as was proposed by S2006 in the case of Z~And. For a mass-transfer rate $10^{-6}~\Msun$~yr$^{-1}$, we found that the mass of unburnt hydrogen accreted in a single outburst is sufficient to trigger a thermonuclear runaway, but the outbursts would then occur every 1.5 to 4 years, instead of the observed 20 yr.

We found, however, that for  mass-transfer rates of few times $10^{-8}\Msun$~yr$^{-1}$ the ignition mass can be accumulated on the WD's surface within roughly 20 yr. In this case a series of dwarf-nova outburst can trigger the recurrent nova eruptions observed in RS~Oph. This model might explain the pre-outburst luminosity increase observed in the light curves of RS~Oph outbursts \citep{adam11} and explain the varying X-ray flux observed at different epochs of RS~Oph quiescence. A recurrent-nova model not involving dwarf-nova triggers requires a very small disc and does not explain the quiescent and pre-outburst observations. This suggests that RS~Oph outbursts are recurrent `combination nova events'. 

\subsection*{\Zand}

We studied the possibility that, as suggested by S2006, some features of  the complex Z~And light-curve can be explained by dwarf-nova outbursts. We found that for a mass-transfer rate of $10^{-6}\Msun$~yr$^{-1}$ the brightness amplitude of the disc instability outburst  is close to 1 mag in the V band, very close to the amplitude of the 1997 outburst. But the disc instability model totally fails to reproduce the observed shape of this outburst: the rise-time is far too long, and the outburst duration too short. For lower mass-transfer rates, in addition to the wrong shape, the outbursts are much fainter than observed. 

Although we have shown that the 1997 feature was not due to a dwarf-nova outburst, we explored the possibility that the 2000--to--2002 outburst of Z~And was, as suggested by S2006, a combination of a dwarf nova and  nova outburst. We found that a single dwarf nova outburst can enhance nuclear burning on the white dwarf provided the mass-transfer rates are close to $10^{-6}~ {\rm M}_{\odot}~{\rm yr}^{-1}$. Such steady high mass-transfer rates, are most unlikely in this system. Reaching such a value requires a large (factor 10 at least) mass-transfer enhancement, that would in itself generate a significant brightening of the disc. Whether this enhancement leads to a disc instability or not is of little importance, and we therefore conclude that the initial component of the proposed `combination nova' scenario is most likely the abrupt enhancement of the mass-transfer rate. Such an enhancement has been proposed by \citet{bp82} for explaining the outbursts of the symbiotic binary CI~Cyg, and could be attributed for example to the magnetic activity at the surface of the companion as suggested by \citet{Leibowitz2008}.

Dwarf-nova outbursts may play a role in triggering RS~Oph recurrent outbursts but are, most probably, of no importance in the case of Z~And.

\section*{Acknowledgements}
We are grateful to Lars Bildsten, Joanna Miko\l ajewska and Bill Wolf for advice and helpful discussions. We thank the referee Chris Tout for his severe but friendly criticism of the first version of this paper. We acknowledge with thanks the variable star observations from the AAVSO International Database contributed by observers worldwide and used in this research. This research was supported by the National Science Centre, Poland, grants 2013/08/A/ST9/00795, 2013/10/M/ST9/00729 and 2015/19/B/ST9/01099, and in part by the National Science Foundation under Grant No. NSF PHY-1125915 to KITP in Santa Barbara. JPL was supported by a grant from the French Space Agency CNES.

\bibliographystyle{mnras}
\bibliography{RS_ZA_JPL}
\newpage
\appendix
\section{Critical values of effective temperature, column density and accretion rate}
\label{sect:critvalues}

\begin{equation}
\label{eq:teffplus}
T_{\rm eff}^+=7122\,\alpha^{-0.002}m^{0.03}R_{10}^{-0.082}\,\rm K
\end{equation}
\begin{equation}
\label{eq:teffminus}
T_{\rm eff}^-=5665\,\alpha^{-0.005}m^{0.03}R_{10}^{-0.087}\,\rm K
\end{equation}
\begin{equation}
\label{eq:Sigmaplus}
\Sigma^+=7.983\,\alpha^{-0.784}m^{0.374}R_{10}^{1.122}\,\mathrm{g\,cm^{-2}}
\end{equation}
\begin{equation}
\label{eq:Sigmaminus}
\Sigma^-=12.16\,\alpha^{-0.835}m^{0.371}R_{10}^{1.113}\,\mathrm{g\,cm^{-2}}
\end{equation}
\begin{equation}
\label{eq:Mdotplusapp}
\dot M^+=9.16\times 10^{15}\,\alpha^{-0.008}m^{-0.88}R_{10}^{2.67}\,\mathrm{g\,s^{-1}}
\end{equation}
\begin{equation}
\label{eq:Mdotminus}
\dot M^-=3.67\times 10^{15}\,\alpha^{-0.02}m^{-0.88}R_{10}^{2.65}\,\mathrm{g\,s^{-1}}
\end{equation}

The `$+$' and `$-$' superscripts distinguish between the critical values for the upper and lower branch respectively. These fits are on the average accurate to 1.5 per cent ($T_{\rm eff}^+$), 3.5 per cent ($T_{\rm eff}^-$), 6.9 per cent ($\Sigma^+$), and 10.3 per cent ($\Sigma^-$),  for $10^{-2} < R_{10} < 100$ and $10^{-4} < \alpha < 1$ with maximum errors of 10, 12, 40 and 35 percent respectively. 

\bsp	
\label{lastpage}
\end{document}